\documentclass[10pt,journal]{IEEEtran}

\usepackage{graphicx}  
\makeatletter
\newcommand{\removelatexerror}{\let\@latex@error\@gobble}
\makeatother
\let\chapter\section
\usepackage[ruled]{algorithm2e}

\LinesNumbered

\usepackage{amsmath}
\usepackage{amsfonts}

\usepackage{bm}

\usepackage{latexsym}
\usepackage{amssymb}

\usepackage{multirow}
\usepackage{booktabs}

\usepackage{stfloats}
\usepackage{color}
\definecolor{ggreen}{rgb}{0.00,0.59,0.00}
\definecolor{bbrown}{rgb}{0.50,0.00,0.50}
\usepackage{caption}

\usepackage{array}
\usepackage{arydshln}
\usepackage{blkarray}
\usepackage{pifont}

\usepackage{subfloat}
\usepackage{subfigure}
\usepackage{url}
\usepackage[numbers,sort&compress]{natbib}

\newtheorem{theorem}{Theorem}[section]

\newtheorem{proposition}[theorem]{Proposition}

\newtheorem{definition}{Definition}[section]

\newtheorem{remark}{Remark}[section]

\renewcommand{\baselinestretch}{0.86}

\def\BibTeX{{\rm B\kern-.05em{\sc i\kern-.025em b}\kern-.08em
    T\kern-.1667em\lower.7ex\hbox{E}\kern-.125emX}}
\ifCLASSINFOpdf
\else
\fi
\begin{document}
%
\title{Analysis of Moving Target Defense Against False Data Injection Attacks on Power Grid}

\author{Zhenyong~Zhang,
        Ruilong~Deng, \emph{Member, IEEE,}
        David~K.~Y.~Yau, \emph{Senior Member, IEEE}\\
        Peng~Cheng, \emph{Member, IEEE,}
        and~Jiming~Chen, \emph{Fellow, IEEE}
\thanks{Manuscript received January 7, 2019; revised April 27, 2019; accepted July 12, 2019. This work was supported in part by the National Key Research and Development Program of China under Grant 2016YFB0800204, National Natural Science Foundation of China under Grant 61833015, Singapore University of Technology and Design - Zhejiang University Innovation, Design and Entrepreneurship Alliance (SUTD-ZJU IDEA) Award No. 201805, Nanyang Technological University (NTU) Internal Funding - Start-up Grant (SUG) - the college of engineering (CoE) (M4082287), and A*STAR-NTU-SUTD AI Partnership Grant (RGANS1906). The associate editor coordinating the review of this manuscript and approving it for publication was Dr. Walid Saad. (\emph{Corresponding author: Peng Cheng.})}
\thanks{Z. Zhang, P. Cheng and J. Chen are with the State Key Laboratory of Industrial Control Technology and the College of Control Science and Engineering, Zhejiang University, Hangzhou 310027, China (e-mail: zhangzhenyong@zju.edu.cn; saodiseng@gmail.com; cjm@zju.edu.cn).}
\thanks{R. Deng is with the School of Computer Science and Engineering, Nanyang Technological University, Singapore 639798 (e-mail: rldeng@ntu.edu.sg).}
\thanks{David~K.~Y.~Yau is with the Information Systems Technology and Design, Singapore University of Technology and Design, Singapore 487372 (e-mail: david$\_$yau@sutd.edu.sg).}
\thanks{A preliminary version of this paper was presented at IEEE PES ISGT 2019. \cite{zhang2019completefdi}}}

\maketitle

\begin{abstract}
Recent studies have considered thwarting false data injection (FDI) attacks against state estimation in power grids by proactively perturbing branch susceptances. This approach is known as moving target defense (MTD). However, despite of the deployment of MTD, it is still possible for the attacker to launch stealthy FDI attacks generated with former branch susceptances. In this paper, we prove that, an MTD has the capability to thwart all FDI attacks constructed with former branch susceptances only if (i) the number of branches $l$ in the power system is not less than twice that of the system states $n$ (i.e., $l \geq 2n$, where $n + 1$ is the number of buses); (ii) the susceptances of more than $n$ branches, which cover all buses, are perturbed. Moreover, we prove that the state variable of a bus that is only connected by a single branch (no matter it is perturbed or not) can always be modified by the attacker. Nevertheless, in order to reduce the attack opportunities of potential attackers, we first exploit the impact of the susceptance perturbation magnitude on the dimension of the \emph{stealthy attack space}, in which the attack vector is constructed with former branch susceptances. Then, we propose that, by perturbing an appropriate set of branches, we can minimize the dimension of the \emph{stealthy attack space} and maximize the number of covered buses. Besides, we consider the increasing operation cost caused by the activation of MTD. Finally, we conduct extensive simulations to illustrate our findings with IEEE standard test power systems.
\end{abstract}


%
\IEEEpeerreviewmaketitle

\section{Introduction}\label{section:introduction}
Modern power grid is becoming more scalable for new devices, more efficient for productions and smarter for operations. For the purpose to realize intelligent automation at all operation levels, numerous sensors and meters are distributed in this large-scale system for wide area monitoring, protection and control. However, these advanced information and communication technology (ICT) components make the power system prone to cyber attacks. It has been proved that the attacker can tamper with measurements in the field devices such as the smart meter \cite{illera2014smartmetersnn} and the remote terminal unit (RTU) \cite{konstantinou2016realfdi}. With increased connectivity of physical power grids to open systems such as the Internet (e.g., convergence between IT and operation technology or OT), it is imperative to enhance the security of power grids to keep intruders at bay \cite{gungor2011smartgrids, kurt2018distributed, kurt2019distributed}.

Traditional supervisory control and data acquisition (SCADA) systems for power grids implement basic integrity and availability checks (e.g., bad data detection or BDD) for their data, to reject erroneous measurements due to failures or malicious attacks such as false data injection (FDI). However, research has shown that carefully designed FDI attacks can bypass the BDD and remain stealthy, when attackers utilize comprehensive knowledge of the system topology and branch susceptances of the power network to guide their actions \cite{liu2011false}. Although stealthy, these FDI attacks \cite{deng2017ccpa, tan2016optimalattack, yuan2012quantiveattak, zhang2017stealthycsa, liu2019attackmicro} can be quite powerful. They may lead to large errors in the estimated system states and cause severe consequences such as prolonged interruption of power supply or equipment damage \cite{giannini2014consequence}.

Addressing the imminent threats posed by stealthy FDI attacks, many recent research efforts have sought to characterize their properties and propose countermeasures against them \cite{deng2018sparsefdi,deng2017fdisurvey,tan2017mitigate}. For example, methods to secure meter measurements and critical state variables against tampering have been proposed~\cite{yang2017pmudeployment} \cite{kim2011strategic}. In practice, however, breach of the perimeter, including cryptographical safeguards, has been repeatedly demonstrated in the real world through persistent attempts by malicious attackers \cite{pal2018pmumanipulationattack, konstantinou2016realfdi, yao2017networktopology, illera2014smartmetersnn}. Besides, since only partial of the measurements are trusted, this strategy may reduce the redundancy of the original monitoring system.

Observing that the construction of stealthy FDI attacks depends on the detailed knowledge of the power grid's configuration, an alternative defense approach is to change the system parameters by design for defeating the knowledgeable attacker. Existing work has typically implemented such {\em moving target defense} (MTD) \cite{Rahaman2014mtd} by proactively perturbing the impedance of certain branches of a power network using the distributed flexible AC transmission system (D-FACTS) devices, e.g., DSI and DSSC \cite{divan2007dfacts}. By modifying D-FACTS, the changes of branch parameters are unpredictable to attackers, thus, increasing uncertainties for them to execute stealthy FDI attacks on the power system, which complies with the definition of MTD.

Prior works on MTD against stealthy FDI attacks have demonstrated success in defeating knowledgeable attackers. Morrow \emph{et al.} \cite{morrow2012perturbation1} and Davis \emph{et al.} \cite{morrow2012perturbation2} investigated the divergence of the system state due to bad/malicious data, by comparing expected results caused by branch impedance perturbations with actual observed responses of the system.
Rahman \emph{et al.} \cite{Rahaman2014mtd} presented the formal design of an MTD application, and show by simulations that arbitrary branch susceptance perturbations may not be effective in detecting FDI attacks. Tian \emph{et al.} \cite{tian2018hidden} proposed a notion of hidden MTD, which aims to make the defense stealthy to attackers. Liu \emph{et al.} \cite{liu2018hmtddistribution} extended the hidden MTD to the AC distribution system, and remarkably, considered the minimization of power losses and power flow differences before and after MTD.
Lakshminarayana and Yau \cite{lakshminarayana2018cost} presented analytical conditions for MTD to be truly effective, and presented an explicit cost-benefit analyis of the MTD, which can be viewed as a form of insurance. Liu \emph{et al.} \cite{liu2018vullr} considered the utility of MTD as an optimized goal, and solved a joint optimization problem with the generation cost loss. Although the authors in \cite{tian2018hidden, lakshminarayana2018cost, liu2018vullr} have analyzed MTD's capability to thwart FDI attacks constructed with former branch parameters, they haven't given insight about the effectiveness of MTD and presented limitations of MTD related to both the power network structure and perturbed branches. Besides, the impact of the perturbation magnitude on the stealthy FDI attack space after MTD is not thoroughly investigated.

In fact, for understanding the impacts of the deployment of D-FACTS devices on the power system, pioneer works have devoted to studying the linear sensitivities of power system quantities such as voltage and power losses with respect to the branch impedance perturbation \cite{lee2001hybridets, rogers2008mtdmag,brissette2014dfacts}. For example, Rogers and Overbye \cite{rogers2008mtdmag} analyzed the linear sensitivities with respect to branch impedance for solving the real power loss minimization and voltage control problems. They showed that, by perturbing the branch impedances of 5 branches within the range +/- 20\% of their original values in the IEEE 14-bus power system with D-FACTS devices, the power loss is 3.35 MW compared to 3.51 MW without any perturbation. Significantly, Morrow \emph{et al.} \cite{morrow2012perturbation1} investigated the impacts of the use of D-FACTS on the power losses and voltages considering the defense effect of branch perturbations. They proved that if the perturbations are within 20\% of the impedance in the IEEE 14-bus power system, then there are nearly 10,000 perturbation cases that can restrict the power losses within 1\%. That is, the operating points after branch perturbations are limited to the small neighborhood around the optimum operation points. Other power systems were also tested. The results highlight the practicality of modifying D-FACTS to provide MTD in power grid.

In this paper, we focus on analyzing the completeness, deployment and the increasing operation cost of MTD in terms of thwarting FDI attacks constructed with old system information. Similar to prior studies \cite{lakshminarayana2018cost,tian2018hidden,Rahaman2014mtd, liu2018vullr}, we mainly consider FDI attacks of the form $\textbf{a} = \mathbf{H}\textbf{c}$  \cite{liu2011false} with the DC power flow model. To begin with, we define a \emph{stealthy attack space} as the intersection of the set of attack vectors generated with the former branch susceptances and the set of attack vectors generated with the current branch susceptances after MTD. Once an MTD is able to reduce the dimension of the \emph{stealthy attack space} to 0, it is defined as a complete MTD. Based on these definitions, we analyze the conditions required for a complete MTD from the aspect of inherent power network structure and the branches that are perturbed. Besides, limitations for a power system to satisfy these conditions are also exploited. Further, for the case when the necessary conditions for a complete MTD are not met, we investigate methods to narrow down the attack opportunities of potential attackers by properly selecting a set of target-perturbation branches\footnote{The target-perturbation branch means that the susceptance of the branch will be perturbed.}. Moreover, we discuss the reduction of the additional operation cost caused by the activation of MTD. It should be clarified that some results have been presented in our conference paper \cite{zhang2019completefdi}, in which we have analyzed the topology limitation for achieving a complete MTD and the impacts of the perturbation magnitude on the reduction of the stealthy attack space. While in this journal version, more issues are discussed, including the special power network structure with which we can never achieve complete MTD, the impacts of the perturbation magnitude on the change (increase, invariant and decrease) of the dimension of the stealthy attack space, the optimal deployment of D-FACTS devices and the additional operation cost caused by MTD. We also present more simulations to illustrate and demonstrate our findings. In summary, the contributions are as follows:
\begin{itemize}
  \item First, we prove that an MTD is complete only if (i) the number of branches $l$ is larger than or equal to twice that of the system states $n$ (i.e., $l \geq 2n$, where $n+1$ is the number of system buses), and (ii) the susceptances of more than $n$ branches, which cover all buses, are perturbed. Besides, we prove that we can never realize a complete MTD if the power network contains a bus that is only connected by a single branch. The state variable of this bus can be injected arbitrary bias by the attacker.
  \item Moreover, we observe that, the change of the perturbation magnitude almost does not affect the dimension of the stealthy attack space. Based on this result, we propose an algorithm to compute a feasible set of target-perturbation branches that can minimize the dimension of the stealthy attack space and maximize the number of covered buses. Besides, we discuss the increase of the operation cost after MTD considering the security constraint, which is associated with the susceptance perturbation magnitude.
  \item Finally, we illustrate and demonstrate our results by conducting extensive simulations with the IEEE standard power systems.
\end{itemize}

The remainder of this paper is organized as follows. We introduce the system model and threat model in Section \ref{section:systemmodel}. The problem statement is addressed in Section \ref{section:problemsetuped}. In Section \ref{section:performanceanalysis}, we analyze the MTD in terms of thwarting FDI attacks. We present simulation results of our findings in Section \ref{section:simulation}. Section \ref{section:conclusion} concludes the paper.

\section{System model and threat model}\label{section:systemmodel}
Throughout this paper, calligraphy font ($\mathcal{L}$) indicates a set or a subspace ($\mathcal{A}$), math boldface font ($\mathbf{H}$) indicates a matrix, bold lower case letter ($\textbf{x}$) indicates an vector. $S(\cdot)$ denotes the (column) span of a matrix. $R(\cdot)$ is the rank of a matrix. $\cdot^T$ indicates the transpose of a matrix. All proofs in this paper are included in the Appendix. $\left |\cdot \right |$ denotes the cardinality of a set.

\subsection{System model}\label{section:systemmodel}
To avoid intensive computation and obtain an optimal solution for large power systems, power system engineers usually utilize a linearized DC power flow model to approximate the AC power flow model \cite{wood1996system}\cite{abur2004stateestimation}. For its computational speed and simplicity, the DC model has been widely used for decades in both industry and academia \cite{awad2010industryuse,geenrgy2007software,stott2009dcpowerflow,PLEXOS2019software}. Although it is less accurate, the DC model is more robust and often used in real-time operations such as the computation of marginal price \cite{li2010marginalprice}. In the DC model, the voltage magnitude is by default set as 1 p.u.. The state variables are reduced to the voltage phase angle. The measurements are reduced to active power flows. Moreover, since the phase angle difference is usually constrained to be small, the power flow equations can be reduced to
\begin{equation}\label{dc}
   f_{ij}= -b_{ij}(\theta_i -\theta_j), p_i = \sum_{j \in \mathcal{K}_i} f_{ij},
\end{equation}
where $f_{ij}$ indicates the active power flow between bus $i$ and bus $j$, $b_{ij} $ is the equivalent susceptance on branch $\{i, j\}$, $\theta_i$ and $\theta_j$ are respectively the voltage phase angle of bus $i$ and $j$, $p_i$ is the active power injection of bus $i$, $\mathcal{K}_i$ is a set of neighboring buses connected to bus $i$.

We consider a classic power transmission network consisting of a set $\mathcal{N}=\{1,2,\cdots, n+1\}$ of buses and a set $\mathcal{L}=\{k_1, k_2,\cdots, k_l\}$ of branches, where $n+1$ is the number of buses and $l$ is the number of branches. With the DC model, by setting an arbitrary bus as the reference/slack bus, the remaining $n$ phase angles $\theta_1, \theta_2, \cdots, \theta_n$ are taken to form the system states, typically denoted as $\textbf{x} \in \mathbb{R}^n$. Each branch $k_t = \{i,j\}\in \mathcal{L}$ connects two buses $i$ and $j$.
Assuming that the power system is fully measured, i.e., each bus is monitored by one meter and each branch is monitored by two meters (both in the positive and negative direction). Then, the number of meter measurements is $m = 2l + n + 1$. The DC model can be derived as
\begin{equation}\label{dcpowerflowmodel}
  \textbf{z} = \mathbf{H}\textbf{x} + \bm{\eta}.
\end{equation}
where $\mathbf{H}$ is termed as the measurement matrix, $\textbf{z}$ denotes the measurements of active power injections and active power flows, $\bm{\eta}$ represents the independent measurement noises, which are typically assumed to be normally distributed [i.e., $\eta_i \sim \mathcal{N}(0, \sigma^2_i)$].

\textbf{Construction of the measurement matrix.} Let $\mathbf{A}$ denote the branch-bus incidence matrix. It is given by
\begin{equation}\label{e2}
  a_{ti}= \begin{cases}
    1, & \text{if branch $k_t$ starts from bus $i$};\\
    -1, & \text{if branch $k_t$ ends at bus $i$}; \\
    0, & \text{otherwise}, \\
  \end{cases}
\end{equation}
where $a_{ti}$ is the element in the position $(t,i)$ of the matrix $\mathbf{A}$. Let $\mathbf{D}$ denote the diagonal branch susceptance matrix. Its diagonal element $\mathbf{D}_{tt}$ is $-b_{ij}$ with branch $k_t = \{i, j\}$. Thus, the invertible symmetric admittance matrix is $\mathbf{B}=\mathbf{A}^T \mathbf{D}\mathbf{A}$ and the branch-bus shift factor matrix is $\mathbf{S} = \mathbf{D}\mathbf{A}$. Considering the DC power flow equations, we have $\textbf{f} = \mathbf{S} \bm{\theta}$ and $\textbf{p}=\mathbf{B}\bm{\theta}$, where $\textbf{f}$ denotes the active power flows, $\textbf{p}$ denotes the power injections, $\bm{\theta}$ denotes the phase angles. Therefore, the measurement matrix $\mathbf{H} \in \mathbb{R}^{m \times n}$ can be derived as
\begin{equation}\label{e4}
  \mathbf{H} = \begin{bmatrix} \mathbf{B}  \\ \mathbf{S} \\ -\mathbf{S} \end{bmatrix} = \begin{bmatrix} \mathbf{A}^T\mathbf{D}\mathbf{A}  \\ \mathbf{D}\mathbf{A} \\ -\mathbf{D}\mathbf{A} \end{bmatrix},
\end{equation}
We can see that the measurement matrix $\mathbf{H} \in \mathbb{R}^{m \times n}$ is a function of the system topology and branch susceptances.

In cases that the power systems are partially measured (i.e., some buses and branches are not monitored by meters), the corresponding measurement matrix is formed by selecting some rows from the measurement matrix of the fully measured case. For any measurement matrix, we assume that $m > n$ in this paper. Note that all results in this paper are satisfied whenever the system is fully measured or partially measured.

\textbf{State estimation (SE).} State estimator, a fundamental tool for economically and dynamically routing power flows, is responsible to optimally estimate the state variables with noisy measurements collected by the underlying SCADA system \cite{wood1996system}. The most common and concise mathematical description of SE is
\begin{equation}\label{estimator}
 \hat{\textbf{x}} = (\mathbf{H}^T\mathbf{W}\mathbf{H})^{-1}\mathbf{H}^T\mathbf{W}\textbf{z} = \mathbf{K}\textbf{z},
\end{equation}
where $\mathbf{W}$ is a diagonal matrix whose elements are reciprocals of the noise variances, $\mathbf{K}$ is referred to as the ``pseudo-inverse" of $\mathbf{H}$ because $\mathbf{K}\mathbf{H}=\mathbf{I}$. $ \hat{\textbf{x}}$ is calculated by using certain statistical estimation criterions such as maximum likelihood, weighted least-squares and etc.

\textbf{Bad data detection (BDD).} Normally, in order to filter out abnormal/erronuous meter measurements, the bad data detection (BDD) checker is used in SE. Let $\textbf{r} = \textbf{z} - \mathbf{H}\hat{\textbf{x}} = (\mathbf{I} - \mathbf{H} \mathbf{K})\textbf{z}$ be the residues between the measurements $\textbf{z}$ and their estimates $\hat{\textbf{z}} = \mathbf{H}\hat{\textbf{x}} $. BDD compares the Euclidean norm $\|\textbf{r}\|_2$ with a predetermined threshold $\tau$. If $\|\textbf{r}\|_2 > \tau$, the abnormal alarm is triggered; otherwise, the measurements $\textbf{z}$ are considered normal.

In the following, we assume that the measurements are noiseless (i.e., $\textbf{e} = \textbf{0}$) to simplify the discussions. Under this assumption, the DC model can be written as $\textbf{z} = \mathbf{H} \textbf{x}$. The corresponding state estimates are given by $\hat{\textbf{x}} = (\mathbf{H}^T\mathbf{H})^{-1}\mathbf{H}^T\textbf{z} = \textbf{x}$. Moreover, the threshold $\tau$ for the BDD is equal to 0. Nevertheless, the main results in the following derivations still keep valid with noisy measurements. In our simulations (see Section \ref{section:simulation}), we will evaluate the impacts of measurement noises.

\subsection{Threat model}\label{section:threatmodel}
As a critical infrastructure, the security issue of power grid has attracted a lot of attention. For example, US National Electric Sector Cybersecurity Organization Resource (NESCOR) records safety incidents and negative impacts on the physical objects in power systems \cite{nescor2019attackinstance}. North American Electrical Reliability Council (NERC) reports lessons learned from failures and blackouts caused by cyber system faults \cite{nerc2019attackinstance}. These recordings show the security risks of power systems, which should be well addressed during daily maintenance and operation.

\textbf{Adversarial setting.} In this paper, we consider the class of stealthy FDI attacks studied in \cite{liu2011false}. Let $\textbf{a} \in \mathbb{R}^m $ be the attack vector. The malicious measurements are given by $\textbf{z}_a = \textbf{z} + \textbf{a}$. It has been proved that $\textbf{z}_a$ cannot be detected by the BDD if $\textbf{a} = \mathbf{H}\textbf{c}$ \cite{liu2011false}, where $\textbf{c} \in \mathbb{R}^n$ is an arbitrary vector. That is,   $\textbf{r}^a = \textbf{z}^a - \mathbf{H}\hat{\textbf{x}}^a = \textbf{z} + \textbf{a} - \mathbf{H}(\hat{\textbf{x}} + \textbf{c})= \textbf{z}- \mathbf{H} \hat{\textbf{x}} = \textbf{r}$,
where $\textbf{r}^a$ denotes the measurement residues after the FDI attack and $\hat{\textbf{x}}^a$ denotes the modified states. We can see that the measurement residues are not changed after the FDI attack. Thus, the bad data alarm is not triggered while the system states are subverted.

In other words, let $\mathcal{A} \in \mathbb{R}^m$ be a subspace that contains all FDI attacks with the form of $\textbf{a} = \mathbf{H}\textbf{c}$. Then, $\mathcal{A}$ is given by
\begin{equation}\label{attackvectorset}
  \mathcal{A} = \{\textbf{a}\mid \textbf{a} = \mathbf{H}\textbf{c}, ~~\textbf{c} \in \mathbb{R}^n\}.
\end{equation}
We can see that $\mathcal{A}$ is equal to the subspace $S(\mathbf{H})$, where $S(\cdot)$ denotes the (column) span of a matrix. Therefore, if $\textbf{a} \in S(\mathbf{H})$ holds, then the malicious measurements $\textbf{z}_a$ can bypass the BDD. In this paper, we assume the attacker has the following capabilities:
\begin{itemize}
  \item The attacker is able to know a measurement matrix $\mathbf{H}$ based on his/her understanding, e.g., through topology-leaking attacks \cite{markwood2017tla} or subspace attacks \cite{kim2015subspaceattack}. Since these topology attacks depend on historical measurements, the measurement matrix $\mathbf{H}$ cannot be learned by the attacker immediately. The learning/inferring process usually takes a sufficiently long time (hours or days) due to the exfiltration of an enough amount of historical measurement data \cite{lakshminarayana2018cost, kim2015subspaceattack, yu2015pca, chen2019learningbased, laks2018datadriven}.
  \item The attacker is able to eavesdrop and tamper with the measurements by intruding into the communication network or IP-accessible field devices \cite{stephen2010penetration}. Practically, the attacker will have limited resources to compromise all the measurements \cite{liu2011false}. However, we do not assume a priori what specific data can be compromised or not. Note also that although the attacker might be able to compromise the confidentiality of the raw data points, he/she will need non-trivial efforts and time to learn their higher system level relationships to guide his attack.
  \item The attacker is unable to take over (or have full access to) the control center or the SCADA systems. Once the attacker can exploit the control center or the SCADA systems, he/she is powerful enough to thwart most defensive mechanisms.
\end{itemize}

\section{Problem Statement}\label{section:problemsetuped}
Based on the system model and threat model, in this section, we first introduce the approach of MTD used in power grid. Then, we present the problems mainly investigated in this paper.

\subsection{Moving target defense by perturbing branch susceptances}\label{section:mtdinpg}
\textbf{D-FACTS.} The distributed flexible AC transmission system (D-FACTS) devices are first introduced by Divan and Johal \cite{divan2007dfacts} for controlling the power flow. These devices can alter the impedances of power lines, and thus, control power flows to eliminate transmission constraints and bottlenecks. They are small and light enough to be suspended from the power line, floating both electrically and mechanically on it. Moreover, the equipped communication system enables them to receive control commands and transmit working states to remote control stations \cite{johal2007designdfacts}. To date, a lot of researches have devoted to analyzing the performance of D-FACTS devices and investigating the use of them in different power system applications \cite{rogers2008mtdmag, brissette2014dfacts, urquizo2017dea, mehta2015defacts, hamidi2017dfactcoperation}.

\textbf{MTD.} Recently, some pioneer works have exploited the adoption of D-FACTS devices for thwarting FDI attacks in power grid \cite{kuntz2015defacts,salehghaffari2018defacts, morrow2012perturbation1, morrow2012perturbation2, Rahaman2014mtd,tian2018hidden,lakshminarayana2018cost,libeibei2019dfacts}. Morrow \emph{et al.} \cite{morrow2012perturbation1} \cite{morrow2012perturbation2} were the first researchers who proposed to perturb branch impedances for probing both the malicious and bad data in the power grid. The following works \cite{Rahaman2014mtd,tian2018hidden,lakshminarayana2018cost} named this branch perturbation strategy as moving target defense (MTD). Here we briefly introduce this defensive mechanism based on the DC model.

With D-FACTS devices, the defender is able to actively perturb a set of branch susceptances, an thus, increase system uncertainty for potential attackers. Supposing the susceptance of branch $k_t = \{i, j\}$ is perturbed, then we have
\begin{equation}\label{reactanceperturbation}
 b_{ij} \rightarrow b'_{ij},
\end{equation}
where $b'_{ij}$ is the susceptance of branch $k_t$ after MTD. Actually, we cannot change $b_{ij}$ as much as we can \cite{rogers2008mtdmag, morrow2012perturbation1, tian2018hidden}. There are limits on the perturbed result, i.e.,
\begin{equation}\label{limitation}
 b^{min}_{ij} \leq b'_{ij} \leq b^{max}_{ij}.
\end{equation}

As a result, the measurement matrix is changed. We assume that the control commands of MTD can be protected in the control center and transmitted through safeguarded communication channels. Thus, the attacker is unable to anticipate them. We use $\mathbf{H}'$ to denote the new measurement matrix after MTD. In fact, the attacker might try to learn the perturbed measurement matrix $\mathbf{H}'$. But the learning process usually takes a sufficiently long time (hours or days) due to the exfiltration of an enough amount of historical measurement data \cite{lakshminarayana2018cost, kim2015subspaceattack, yu2015pca, chen2019learningbased, laks2018datadriven}. In other words, the attacker cannot obtain the latest measurement matrix immediately. Thus, we can dynamically change the measurement matrix accordingly before it risks being exposed. The execution cycle of MTD has been discussed by authors in \cite{tian2018hidden} and \cite{lakshminarayana2018cost}.

Since the current susceptance perturbations are not anticipated by the attacker, he/she only knows the former measurement matrix $\mathbf{H}$ (may not be the measurement matrix just before MTD). Thus, if the attacker still constructs the attack vector as $\textbf{a} = \mathbf{H}\textbf{c}$, it is possible that $\textbf{a} \notin S(\mathbf{H}')$ after MTD. In other words, the malicious measurements $\textbf{z}_a = \textbf{z}' + \mathbf{H}\textbf{c}$ may not bypass the BDD with $\mathbf{H}'$, where $\textbf{z}'$ are the measurements after MTD. The defender's BDD after MTD is given by
\begin{equation}\label{e5}
  r(\textbf{z}') = \|\textbf{z}'-\mathbf{H}'\mathbf{K}'\textbf{z}'\|,
\end{equation}
where $\mathbf{K}' =({\mathbf{H}'}^T\mathbf{H}')^{-1}{\mathbf{H}'}^T$. Since $m>n$, we can prove that $\mathbf{H}'\mathbf{K}'\neq \mathbf{I}$. Therefore, $r(\textbf{z}') = 0$ if and only if $\textbf{z}' \in S(\mathbf{H}')$ with noiseless measurements \cite{liu2011false}. Since $\textbf{z}' \in S(\mathbf{H}')$, $r(\textbf{z}_a) = 0$ if and only if $\textbf{a} \in S(\mathbf{H}')$. This means that attack vector can bypass the BDD after MTD if and only if it belongs to the following \emph{stealthy attack space}.
\begin{definition}
\emph{Let} $\mathcal{A}_s$ \emph{denote the stealthy attack space. Then,} $\mathcal{A}_s = S(\mathbf{H}) \cap S(\mathbf{H}')$\emph{, i.e.,}
\begin{equation}\label{attackspace}
  \mathcal{A}_s = \{\textbf{a}\mid \textbf{a} = \mathbf{H}\textbf{c}~\land~ \textbf{a} = \mathbf{H}'\textbf{c}' , ~~\textbf{c}, \textbf{c}' \in \mathbb{R}^n \}.
\end{equation}
\end{definition}
We can see that the stealthy attack space is the intersection of the column space of $\mathbf{H}$ and $\mathbf{H}'$. Therefore, the stealthy attack space is also a subspace and its dimension is given by
\begin{equation}\label{zerorank}
\begin{aligned}
    &dim(\mathcal{A}_s)\\ &= dim(S(\mathbf{H})\cap S(\mathbf{H}'))\\ & =  dim(S(\mathbf{H})) + dim(S(\mathbf{H}')) - dim\big(S(\mathbf{H}) \cup S(\mathbf{H}')\big)\\
  & =  2n - R([\mathbf{H}~\mathbf{H}']),
\end{aligned}
\end{equation}
where $dim(\cdot)$ is the dimension of a subspace, $R(\cdot)$ is the rank of a matrix. We can see that the dimension of the stealthy attack space $\mathcal{A}_s$ is closely related to the rank of the combined matrix $[\mathbf{H}~\mathbf{H}']$, i.e., $R([\mathbf{H}~\mathbf{H}'])$. We define $R([\mathbf{H}~\mathbf{H}'])$ as the security factor of MTD and denote it as $\gamma$.


\begin{figure*}[htbp]
\begin{minipage}[t]{0.22\linewidth}
\centering
\includegraphics[width=2.5cm]{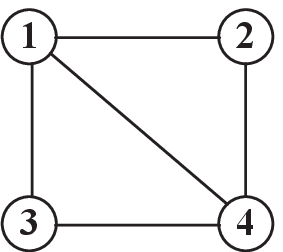}
\caption{ A 4-bus power system with 5 branches.}\label{fig:4-bus-power-system}
\end{minipage}%
\hfill
\hfill
\begin{minipage}[t]{0.54\linewidth}
\centering
\includegraphics[width=7.5cm]{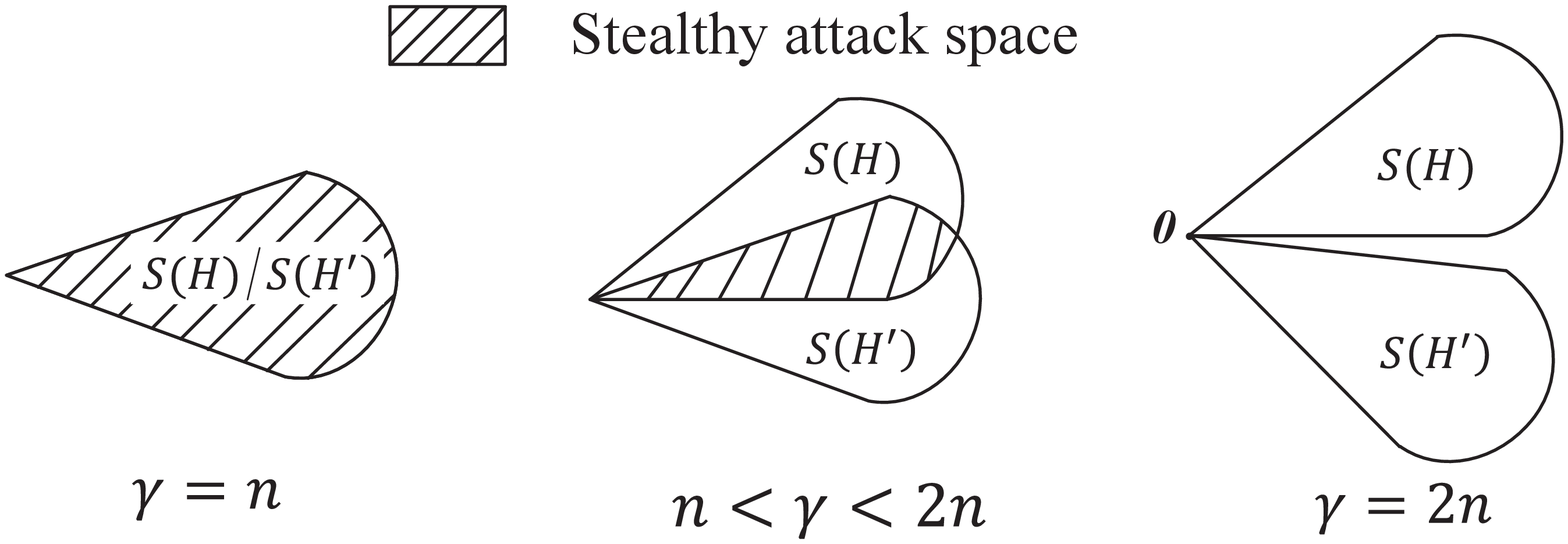}
\caption{An illustration example about the change of the stealthy attack space $\mathcal{A}_s$ with the security factor $\gamma$.}\label{fig:intersection_new_all_correct_representation}
 \end{minipage}%
\hfill
\hfill
\begin{minipage}[t]{0.22\linewidth}
\centering
\includegraphics[width=2.5cm]{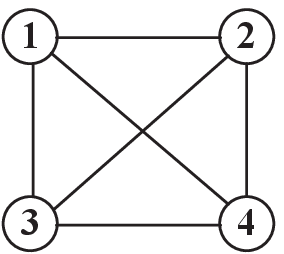}
\caption{ A 4-bus power system with 6 branches.}\label{fig:4bus6branches}
\end{minipage}%
\hfill
\vspace{-0.2cm}
\end{figure*}


\subsection{Problem statement}\label{section:problemsetup}
In this paper, we mainly analyze the effectiveness of MTD from two aspects: first, the capability of MTD to protect the system from any FDI attack constructed with the former measurement matrix; second, if the first capability cannot be achieved, the capability of MTD to narrow the attack opportunities of potential attackers by appropriately deploying the D-FACTS devices and setting the susceptance perturbation magnitudes.

As we know, as long as the stealthy attack space $\mathcal{A}_s$ is not $\{\textbf{0}\}$, it is still possible for the attacker to successfully execute FDI attacks after MTD. The intuitive understanding is that, if $\exists ~ \textbf{c}, \textbf{c}' \in \mathbb{R}^{n}$ satisfying $\textbf{a} = \mathbf{H} \textbf{c} = \mathbf{H}' \textbf{c}'\neq \textbf{0}$ [i.e., $\textbf{a} \in S(\mathbf{H}) \cap S (\mathbf{H}')$], then the malicious measurements $\textbf{z}_a = \textbf{z}' + \textbf{a}$ after MTD can circumvent the defender's BDD. We take the 4-bus power system (Fig. \ref{fig:4-bus-power-system}) as an example. Suppose that the branch susceptances known by the attacker are $b_{12} = -1$, $b_{13} = -2$, $b_{14} = -3$, $b_{23} = -4$ and $b_{34} = -5$.
If we perturb the branches $\{1,2\}$ and $\{2,3\}$ as $\Delta b_{12} = 0.1$ and $\Delta b_{13} = 0.2$, the attacker still can construct an attack vector as $\textbf{a} = [0, 0, 3c, 0, 5c]^T$ to bypass the BDD after MTD, where $c \in \mathbb{R}$ is the error injected into the state variable of bus 4. Here the dimension of the stealthy attack space is 1. In other words, the MTD may not be complete to thwart all FDI attacks with the form of $\textbf{a} = \mathbf{H}\textbf{c}$. We define the MTD's completeness as follows:
\begin{definition}\label{definition:completemtd}
\emph{An MTD is said to be complete if and only if all attack vectors in} $\mathcal{A}$ \emph{except the zero vector cannot bypass the BDD after MTD. This indicates that} $\mathcal{A}_s = \{\textbf{0}\}$.
\end{definition}

Besides, once a complete MTD cannot be achieved, we need to exploit how to reduce the stealthy attack space as much as possible, thus, reducing the probability of stealthy FDI attacks after MTD. An intuitive idea is to perturb all branches in the power network. But it is an unrealistic assumption to deploy D-FACTS on all branches. On the other hand, we cannot randomly choose a set of target-perturbation branches and arbitrarily perturbed them. The impacts on the stealthy attack space might be different if we perturb different branches and set different perturbation values. For example, given the 4-bus power system shown in Fig. \ref{fig:4-bus-power-system}, Table \ref{table:perturbationdifferentbranches} shows the change of the dimension of the attack space when we perturb different branches and set different perturbation values. From row 1 and row 2, we can see that the dimensions of the stealthy attack space are different when we perturb different branches. Form row 3 and row 4, we can see that the dimensions of the stealthy attack space are different when we perturb branch $\{3,4\}$ to different values.
\begin{table}[htbp]
\centering
 \caption{The dimension of the stealthy attack space after MTD}\label{table:perturbationdifferentbranches}
 \begin{tabular}{c|c}
  \hline
  Branch Perturbations&  $dim(\mathcal{A}_s)$\\
  \hline
  $\Delta b_{12} = 0.1, \Delta b_{13} = 0.1$& 1 \\
   \hline
  $\Delta b_{12} = 0.1, \Delta b_{23} = 0.1$ & 2 \\
   \hline
  $\Delta b_{13} = 0.1, \Delta b_{14} = 0.3, \Delta b_{34} = 0.5$ & 2  \\
     \hline
  $\Delta b_{13} = 0.1, \Delta b_{14} = 0.3, \Delta b_{34} = 0.2$ & 1  \\
   \hline
 \end{tabular}
 \vspace{-0.2cm}
\end{table}

Moreover, due to the physical limitation of D-FACTS devices \cite{brissette2014dfacts} and their induced operation costs \cite{lakshminarayana2018cost, tian2018hidden}, the branch susceptance cannot be perturbed to any value. Therefore, a natural question emerges that, given a set of D-FACTS devices, how do they affect the stealthy attack space and the operation cost when we deploy them on different branches and set them with different values$?$ This is a critical issue for us to carry out MTD.

\section{Analysis of MTD to thwart FDI attacks}\label{section:performanceanalysis}
In this section, first, we analyze the detection of FDI attacks constructed with the former measurement matrix using MTD. Second, we analyze the completeness of MTD and discuss physical limitations of the power network to achieve this property. Third, we investigate the impact of the susceptance perturbation magnitude on the dimension of the stealthy attack space. Lastly, we provide guidance on effective MTD for minimizing the dimension of the stealthy attack space, maximizing the number of covered buses, and reducing the operation cost.

\subsection{Detecting FDI attacks with MTD}\label{section:detectionandidentification}
First of all, we discuss the detection of FDI attacks constructed with the former measurement matrix.

After MTD, from the measurement matrix $\mathbf{H}$, we select $u$ columns which form a submatrix $\mathbf{H}^u$ such that $\mathbf{H}^u$ can be linearly represented by $\mathbf{H}'$. The rest $v$ $(v= n-u)$ columns in $\mathbf{H}$ form a submatrix $\mathbf{H}^v$, i.e., $\mathbf{H} = [\mathbf{H}^u~\mathbf{H}^v]$. Assume that the attacker constructs an attack vector as $\textbf{a} = \mathbf{H}\textbf{c}$ with the purpose to bypass the BDD after MTD. Actually, $\textbf{a} = \mathbf{H}^u\textbf{c}_u + \mathbf{H}^v\textbf{c}_v$, where $\textbf{c}_u \in \mathbb{R}^u$ and $\textbf{c}_v \in \mathbb{R}^v$. Then, we have the following conclusion.

\begin{proposition}\label{proposition:detectionoffdiandidentify}
\emph{The BDD after MTD can detect the attack vector} $\textbf{a} = \mathbf{H}\textbf{c}$ \emph{only if} $\textbf{c}_v \neq \textbf{0}_{ v}$ \emph{, where} $\textbf{0}_{ v}$\emph{ is a} $v$\emph{-dimension zero vector.}
\end{proposition}

\indent\indent \emph{Proof.} Please see Appendix \ref{section:theoremfindineidne}.

Proposition \ref{proposition:detectionoffdiandidentify} presents a necessary condition for the detection of FDI attacks constructed with the former measurement matrix. It also indicates the soundness of MTD. That is, if the BDD after MTD can detect the attack vector $\textbf{a}$, then at least one of the state variables in $\textbf{c}_v$ is modified. With this condition, if $\mathbf{H}^v$ cannot be linearly represented by $\mathbf{H}'$, then there might not exist $\textbf{c}'\in \mathbb{R}^n$ such that $\mathbf{H}^v\textbf{c}_v = \mathbf{H}'\textbf{c}'$ for an arbitrarily selected $\textbf{c}_v$. Thus, at most $v$ state variables corresponding to $\textbf{c}_v$ cannot be independently modified by the attacker. Since $v \geq n - dim(\mathcal{A}_s)$, if the dimension of the stealthy attack space $dim(\mathcal{A}_s)$ is smaller, the more state variables cannot be independently modified by the attacker after MTD. However, note that the attackers cannot anticipate/predict the new measurement matrix $\mathbf{H}'$, the ``blind'' attacker might try his/her luck to execute FDI attacks, and thus, increase the detection probability of these attacks with the BDD after MTD.

Specifically, to illustrate the effectiveness of MTD for thwarting FDI attacks, we give an illustration example (Fig. \ref{fig:intersection_new_all_correct_representation}) about the change of the stealthy attack space with respect to the security factor $\gamma$. We can see that, the stealthy attack space is equal to the column space of $\mathbf{H}$ and $\mathbf{H}'$ [$S(\mathbf{H}) = S(\mathbf{H}')$] when the security factor $\gamma$ is $n$; the stealthy attack space is the intersection of the column space of $\mathbf{H}$ and $\mathbf{H}'$ [$S(\mathbf{H}) \neq S(\mathbf{H}')$] when the security factor $\gamma$ is between $n$ and $2n$; the stealthy attack space is $\{\textbf{0}\}$ when the security factor $\gamma$ is $2n$. Therefore, if the security factor $\gamma$ is smaller, the ``volume" of the stealthy attack space is smaller. A smaller stealthy attack space means a smaller probability of stealthy FDI attacks constructed with the former measurement matrix. In other words, after MTD, there are less successful opportunities for the attacker to execute stealthy FDI attacks with the old system information. Thus, we claim that MTD is more effective to thwart stealthy FDI attacks. For example, when $\gamma = 2n$, after MTD, no attack vector $\textbf{a} = \mathbf{H}\textbf{c}$ can bypass BDD, which means that the probability of stealthy FDI attacks constructed with the former measurement matrix is zero. We will evaluate the effectiveness of MTD with respect to the dimension of the stealthy attack space in Section \ref{section:effectivenessofmtd}.

Considering the 4-bus power system shown in Fig. \ref{fig:4bus6branches}, the branch susceptances known by the attacker are $b_{12} = -1$, $b_{13} = -2$, $b_{14} = -3$, $b_{23} = -4$, $b_{24} = -5$ and $b_{34} = -6$. If we do not perturb any branches, then all attack vectors $\textbf{a} = \mathbf{H}\textbf{c}$ can bypass the BDD (i.e., $\gamma = n = 3 $, $ dim(\mathcal{A}_s) = 3$); If we perturb the susceptance of branch $\{1,2\}$ to $b'_{12} = -1.1$, then we can reduce the dimension of the stealthy attack space to 2 (i.e., $ 3 < \gamma = 4 < 6$, $ dim(\mathcal{A}_s) = 2$), but there still exist attack vector $\textbf{a} = \mathbf{H}\textbf{c}$ that can bypass the BDD after MTD, e.g., $\textbf{c} = [0, c_2, c_3]^T$, $c_2$, $c_3 \in \mathbb{R}$; If we perturb the susceptances of branches $\{1,2\}$, $\{1,3\}$ and $\{2,4\}$ as $b'_{12} = -1.1$, $b'_{13} = -2.15$, $b'_{24} = -5.1$, then any FDI attack constructed as $\textbf{a} = \mathbf{H}\textbf{c}$ ($ \textbf{a} \neq \textbf{0}$) cannot bypass the BDD after MTD (i.e., $ \gamma = 2n = 6$, $ dim(\mathcal{A}_s) = 0$). That is, this MTD is complete. We find that we can realize a complete MTD only if more than 3 branches are perturbed under this 4-bus power system with 6 branches. The completeness of MTD is investigated in the next subsection.

\subsection{Analysis of MTD's completeness}\label{section:conditionforcompletemtd}
Here we first give a sufficient and necessary condition for achieving a complete MTD mathematically. Then, we present physical constraints for achieving this property.

\subsubsection{Mathematical analysis}\label{section:mathematicalanalysis}
In fact, an MTD is not complete because the BDD after MTD misses detecting some FDI attacks, i.e., $\mathcal{A}_s \neq \{\textbf{0}\}$. According to the Definition \ref{definition:completemtd}, to achieve a complete MTD, we need to make $\mathcal{A}_s = \{\textbf{0}\}$. Since the subspace spanned by the zero vector has zero dimension, the dimension of $\mathcal{A}_s$ is zero if an MTD is complete. Thus, we have the following conclusion.

\begin{proposition}\label{proposition:theoremvaluable}
\emph{An MTD is complete if and only if the security factor }$\gamma= 2n$.
\end{proposition}

\indent\indent \emph{Proof.} Please see Appendix \ref{section:propossitiongwg}.

Proposition \ref{proposition:theoremvaluable} presents a sufficient and necessary condition for achieving a complete MTD mathematically.  Note that if $\gamma = R([\mathbf{H}~\mathbf{H}']) = 2n$, the intersection of $S(\mathbf{H})$ and $S(\mathbf{H}')$ is $\{\textbf{0}\}$. Thus, all FDI attacks constructed as $\textbf{a} = \mathbf{H}\textbf{c}$ ($\textbf{a} \neq \textbf{0}$) can be detected by the BDD after MTD.

\begin{remark}\label{remark:identificationwithcompletemtd}
Moreover, if an MTD is complete, then we can identify the attack vector constructed with the former measurement matrix. Suppose the attack vector is $\textbf{a} = \mathbf{H}\textbf{c}$. Then, the malicious measurements are $\textbf{z}_a = \mathbf{H}'\textbf{x}' + \mathbf{H}\textbf{c} = [\mathbf{H}'~\mathbf{H}] [ \textbf{x}' ~ \textbf{c}]^T$, where $\textbf{x}'$ denotes the state variables without FDI attacks. Since the combined matrix $[\mathbf{H}'~\mathbf{H}]$ has full column rank, we can uniquely solve the variables $\textbf{c}$ with the malicious measurements. Thus, we can identify the injected errors of the state variables.
\end{remark}

\subsubsection{Physical limitations}\label{section:physicallimitation}
Essentially, whether we can construct a complete MTD or not depends on the structure of the power network and the perturbed branches. Considering the fully measured case, the measurement matrix before and after MTD are

\begin{equation}\label{fullymeasuredbeforemtd}
  \mathbf{H} = \begin{bmatrix} \mathbf{A}^T\mathbf{DA} \\ \mathbf{DA} \\  -\mathbf{DA} \end{bmatrix}~~~~~~~\mathbf{H}' = \begin{bmatrix} \mathbf{A}^T\mathbf{D}'\mathbf{A} \\ \mathbf{D}'\mathbf{A} \\ - \mathbf{D}'\mathbf{A} \end{bmatrix}.
\end{equation}

The combined matrix of $\mathbf{H}$  and $\mathbf{H}'$ can be written as $\mathbf{C} = [\mathbf{H}~\mathbf{H}']$. Let $\widetilde{\mathbf{S}} = \begin{bmatrix} \mathbf{S} & \mathbf{S}' \end{bmatrix} = \begin{bmatrix} \mathbf{D}\mathbf{A} & \mathbf{D}'\mathbf{A} \end{bmatrix} $, we have
\begin{equation}\label{fullymeasuredaftermtd}
  \mathbf{C} = \begin{bmatrix} \mathbf{A}^T\widetilde{\mathbf{S}}\\ \widetilde{\mathbf{S}} \\ - \widetilde{\mathbf{S}} \end{bmatrix} = \begin{bmatrix} \mathbf{A}^T & \textbf{0}_{n\times l} & \textbf{0}_{n\times l} \\ \textbf{0}_{l\times l} & \mathbf{I}_{l\times l} & \textbf{0}_{l\times l} \\ \textbf{0}_{l\times l} & \textbf{0}_{l\times l} & -\mathbf{I}_{l\times l} \end{bmatrix} \begin{bmatrix} \widetilde{\mathbf{S}} \\ \widetilde{\mathbf{S}} \\ \widetilde{\mathbf{S}}\end{bmatrix},
\end{equation}
where $\textbf{0}_{n\times l}$ is an $n$ by $l$ zero matrix, $\mathbf{I}_{l\times l}$ is an $l$ by $l$ identity matrix. Therefore, we can derive that $R(\mathbf{C}) \leq R(\widetilde{\mathbf{S}})$. Since $\widetilde{\mathbf{S}}$ is an $l$ by $2n$ matrix, we have $R(\widetilde{\mathbf{S}}) \leq min\{l,2n\}$. Therefore, only if $l \geq 2n$ can we make $R(\mathbf{C}) = 2n$. In other words, in order to make the security factor $\gamma = R(\mathbf{C}) = 2n$, the number of branches of the power transmission system must be larger than or equal to $2n$. We observe that this condition must be satisfied under the other measured cases (i.e., the system is not necessarily fully measured) as well.
\begin{theorem}\label{theorem:conditionforcompletedadsa}
\emph{An MTD is complete only if the following two conditions are satisfied:}
\begin{itemize}
  \item $l \geq 2n$, where $l$ is the number of branches and $n+1$ is the number of buses in the power system;
  \item \emph{The perturbed branches must cover all buses;}
\end{itemize}
\end{theorem}

\indent\indent \emph{Proof.} Please see Appendix \ref{section:theorem43addad}.

Theorem \ref{theorem:conditionforcompletedadsa} provides a necessary condition for realizing a complete MTD. That is, the completeness of MTD can be achieved only if the system topology and the perturbed branches meet certain requirements. Besides, we can derive another two points from the second condition. First, to cover all buses, at least $n$ branches should be perturbed. Second, once a bus is not covered by the perturbed branches, the state variable of this bus can be modified stealthily if the attacker happens to attack it only. That is, let $\textbf{a} = \mathbf{H}\textbf{c}$ be an attack vector. Then, if $c_i = 0$ ($c_i \in \textbf{c}$) for any $i \in \mathcal{M}_{mtd}$, then the malicious measurements $\textbf{z}_a = \textbf{z}' + \textbf{a}$ can bypass the BDD after MTD, where $\mathcal{M}_{mtd}$ is a non-empty set of buses that are covered by the perturbed branches. Thus, $c_i \neq 0$ ($i \notin \mathcal{M}_{mtd}$) can be any values. This indicates that the attacker can arbitrarily modify the state variables of buses that are not covered by the perturbed branches. Therefore, for a complete MTD, we need to ensure that: (i) the power transmission system has more than $2n$ branches; (ii) more than $n$ branches are perturbed; (iii) the perturbed branches cover all buses. For example, the 4-bus power system shown in Fig. \ref{fig:4bus6branches} can support the realization of a complete MTD, because it has 6 branches ($2n = 2*3 = 6$), while the 4-bus power system with 5 branches shown in Fig. \ref{fig:4-bus-power-system} cannot. If any one of the above three conditions is not satisfied, we cannot make an MTD complete.

Considering the first condition, if the power system has less than $2n$ branches, then the dimension of the stealthy attack space satisfies
\begin{equation}\label{dim}
  dim(\mathcal{A}_s) = 2n - R([\mathbf{H}~\mathbf{H}'])\geq 2n - R(\widetilde{\mathbf{S}}) \geq 2n-l.
\end{equation}
That is, the smallest dimension of the stealthy attack space after MTD is larger than or equal to $2n-l$.

\begin{figure}[!htbp]
  \centering
  \subfigure[]{
    \label{fig:threebusnewfour} 
    \includegraphics[width=1.45in]{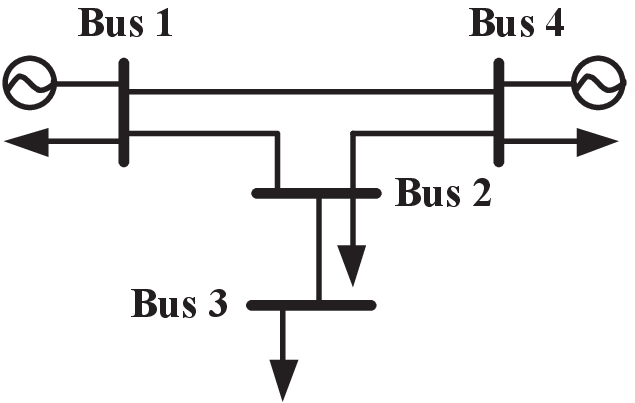}}
  \subfigure[]{
    \label{fig:three-bus_extension} 
    \includegraphics[width=1.45in]{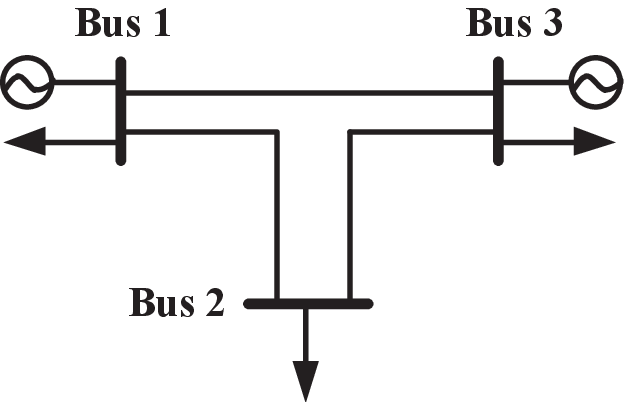}}
  \caption{(a) A 4-bus power system with 4 branches; (b) A 3-bus power system.}
  \label{fig:example} 
\end{figure}

\begin{remark}
\emph{According to the aforementioned analysis, we find that is difficult to realize a complete MTD. For a complete MTD, the power transmission system must have at least} $2n$ \emph{branches (i.e.,} $l \geq 2n$\emph{), and more than} $n$ \emph{branches, which cover all buses, should be perturbed. These conditions may not be held in practice. Particularly, we examine the number of branches in the IEEE test power systems provided in MATPOWER \cite{zimmerman2011matpower}; only three of all 41 cases have more than }$2n$ \emph{branches, namely case6ww (11 branches), case89pegase (210 branches) and case145 (453 branches). What's more, we discover that we can never make an MTD complete if the the power transmission system has a bus that is only connected by a single branch.}

\begin{theorem}\label{theorem:neverscannot}
\emph{It is impossible to make an MTD complete if the power transmission system contains a bus that is only connected by a single branch.}
\end{theorem}

\indent\indent \emph{Proof.} Please see Appendix \ref{section:theorem44naifnsnfi}.

This special case of the power network structure limits the realization of a complete MTD. Besides, we find that the state variable of the bus that is only connected by a single branch can be arbitrarily modified by the attacker. Let $t$ be the bus that is only connected by a single branch. Then, for any $c_t \in \mathbb{R}$, there exists $c'_t \in \mathbb{R}$ such that $\textbf{h}_t c_t = \textbf{h}'_t c'_t$, where $\textbf{h}_t$ and $\textbf{h}'_t$ are respective the $t$th column of the matrix $\mathbf{H}$ and $\mathbf{H}'$. Therefore, if the attack vector is $\textbf{a} = \textbf{h}_t c_t$, then $\textbf{a} \in S(\mathbf{H}')$ always holds, which can definitely bypass the BDD after MTD. We present a simple example to illustrate that. In the 4-bus power system shown in Fig. \ref{fig:threebusnewfour}, the state variable of bus 3 can be arbitrarily modified because it is only connected by branch $\{2,3\}$. Therefore, as long as the system contains a bus that is only connected by a single branch, we can never realize a complete MTD.
\end{remark}

\subsubsection{Discussion}
Nevertheless, it might not be necessary to achieve a complete MTD in some power systems. We take the 3-bus power system shown in Fig. \ref{fig:three-bus_extension} as an example. Even though we cannot realize a complete MTD because it only has 3 branches, which is less than $2*2=4$, we can protect all state variables from being arbitrarily modified by the attacker when we perturb the branches $\{1,2\}$ and $\{2,3\}$. Suppose the perturbations are $\Delta b_{12}$ and $\Delta b_{23}$, and the attack vector is $\textbf{a} = \mathbf{H}\textbf{c}$, where $\textbf{c} = [c_1~c_2]^T$. If this attack can bypass the BDD after MTD, $c_1$ and $c_2$ must satisfy
\begin{equation}\label{attackvectorsatisfication}
  \frac{c_2}{c_1}= \frac{b_{12}\Delta b_{23} - b_{23}\Delta b_{12}}{b_{12} \Delta b_{23} + \Delta b_{12} \Delta b_{23}}.
\end{equation}
Therefore, the attacker must know the susceptance perturbations $\Delta b_{12}$ and $\Delta b_{23}$ to construct a coordinated attack vector $\textbf{a}$. We can see that is almost impossible for the attacker to construct such an attack vector. Definitely, a complete MTD can thwart any attack vector $\textbf{a} = \mathbf{H}\textbf{c}$ when $\mathbf{H}$ is very different from the measurement matrix $\mathbf{H}'$ after MTD. But it might not be a necessary condition for some power systems.

\subsection{Impact of the susceptance perturbation magnitude on the dimension of the stealthy attack space}\label{section:reduceattackspace}
A practical and effective way to enhance the security of the power system is to reduce the stealthy attack space $\mathcal{A}_s$ and cover as many buses as possible with MTD, thus reducing the attack opportunities of potential attackers. We start by investigating the impact of the susceptance perturbation magnitude on the dimension of the stealthy attack space when one more branch susceptance is perturbed. Since the security factor $\gamma$ [i.e., $R([\mathbf{H} ~ \mathbf{H}'])$] determines $dim(\mathcal{A}_s)$, we focus on the change of $\gamma$ in the following. As $\mathbf{H}' = \mathbf{H} + \Delta \mathbf{H}$, based on the sparse property of $\Delta \mathbf{H}$, we first present three cases about the change of $\gamma$ when one more branch is perturbed. Second, we investigate the impact of the susceptance perturbation magnitude on the value of $\gamma$ with these three cases.

Before go deeper into the analysis, we prove that $\Delta \mathbf{H}$ has a sparse structure.
\begin{proposition}\label{proposition:sparsedeltaHdada}
\emph{Suppose} $\mathcal{L}_{d}$ \emph{is a set of perturbed branches.} $\Delta \mathbf{H}$ \emph{is a sparse matrix with non-zero elements in the} $i_d$\emph{th and } $j_d$\emph{th columns, with} $k_d = \{i_d,j_d\} \in \mathcal{L}_{d}$.
\end{proposition}

\indent\indent \emph{Proof.} Please see Appendix \ref{section:propositiondadau}.

Based on the sparse structure of the matrix $\Delta \mathbf{H}$, we can draw a conclusion about the changing range of $\gamma$ when one more branch is perturbed. Let $\mathbf{C} = [\mathbf{H}~\mathbf{H}']$ and $\mathbf{C}' = [\mathbf{H}~\mathbf{H}'']$ be the combined matrices before and after the perturbation of a new branch, respectively. We denote $\Delta \gamma = R(\mathbf{C}') - R(\mathbf{C})$ as the change of the security factor $\gamma$. Then, we obtain the following result.

\begin{proposition}\label{propostion:pppp}
\emph{When perturbing one more branch,} $\Delta \gamma $ \emph{changes -1, 0 or 1.}
\end{proposition}

\indent\indent \emph{Proof.} Please see Appendix \ref{section:propostionsdsfdad}.

That is, there are three possible changes of the security factor $\gamma$ when one more branch is perturbed: increasing 1, remaining the same or decreasing 1 (i.e., $\Delta \gamma = 1, 0$ or $-1$). Thus, a natural question emerges that, whether the value of $\Delta \gamma$ will change with the susceptance perturbation magnitude. Let $k_d = \{i_d, j_d\}$ be the new branch whose susceptance $b_{i_d j_d}$ is perturbed to be $\lambda b_{i_d j_d}$ with $\lambda > 0$ and $\lambda \neq 1$. We define $\lambda$ as the perturbation ratio. Suppose $\mathcal{M}^q_{mtd}$ is an index set of $q$ columns in $\mathbf{H}'$ that form a submatrix $\mathbf{H}^q$ such that $R([\mathbf{H} ~ \mathbf{H}^q]) = R([\mathbf{H}~\mathbf{H}']) = n + q$. The rest columns in $\mathbf{H}'$ form a submatrix $\mathbf{H}^{p}$, i.e., $\mathbf{H}' = [\mathbf{H}^q ~ \mathbf{H}^p]$. We define $\mathcal{M}^q_{mtd}$ as a security set. Overall, there are three cases should be considered for the impacts on $\Delta \gamma$ when perturbing one more branch:
\begin{itemize}
  \item \emph{\textbf{Case 1:}} Neither bus $i_d$ nor $j_d$ are contained in the security set. That is, $i_d \notin \mathcal{M}^q_{mtd}$ and $j_d \notin \mathcal{M}^q_{mtd}$;
  \vspace{2pt}
  \item \emph{\textbf{Case 2:}} Either bus $i_d$ or $j_d$ is contained in the security set. That is, $i_d \in \mathcal{M}^q_{mtd}$ and $j_d \notin \mathcal{M}^q_{mtd}$, or $i_d \notin \mathcal{M}^q_{mtd}$ and $j_d \notin \mathcal{M}^q_{mtd}$;
  \vspace{2pt}
  \item \emph{\textbf{Case 3:}} Both bus $i_d$ and $j_d$ are contained in the security set. That is, $i_d \in \mathcal{M}^q_{mtd}$ and $j_d \in \mathcal{M}^q_{mtd}$.
\end{itemize}

For each case, we analyze the change of $\Delta \gamma$ by varying the perturbation ratio $\lambda$. Since there are limits on the susceptance perturbation \cite{rogers2008mtdmag}, we assume $\lambda_{min}\leq \lambda \leq \lambda_{max}$. We find that the magnitude of the susceptance perturbation almost does not affect the change of the security factor. Especially, if the security factor increases 1 (i.e., $\Delta \gamma = 1$), this result remains the same regardless of the perturbation magnitude. We present the details of the analysis in the following.

\begin{proposition}\label{proposition:proforrankincreasingcase1}
\emph{Under} \emph{\textbf{Case 1}}\emph{, if} $i_d \notin \mathcal{M}^q_{mtd}$ \emph{and} $j_d \notin \mathcal{M}^q_{mtd}$\emph{, then} $\Delta \gamma$ \emph{remains the same regardless of the change of} $\lambda$.
\end{proposition}

\indent\indent \emph{Proof.} Please see Appendix \ref{section:propsoitiond47nins}.

Proposition \ref{proposition:proforrankincreasingcase1} implies that, under \emph{\textbf{Case 1}}, the perturbation magnitude does not affect the change of $dim(\mathcal{A}_s)$. Therefore, if we find that the dimension of the stealthy attack space $dim(\mathcal{A}_s)$ decreses 1 when we perturb a branch to a certain value, then this result will not change if we perturb the branch to the other values. That is, we can determine the value of $dim(\mathcal{A}_s)$ by only testing one perturbation ratio under \emph{\textbf{Case 1}}. Considering the other two cases, we obtain the following result.

\begin{proposition}\label{proposition:decrease}
\emph{Under} \emph{\textbf{Case 2}} \emph{and} \emph{\textbf{Case 3}}\emph{, only if there exists a value} $\lambda^*$ ($\lambda_{min}\leq \lambda^* \leq \lambda_{max}$, $\lambda^* \neq 1$)\emph{ and} $\lambda = \lambda^*$, \emph{we obtain} $\Delta\gamma = -1$.
\end{proposition}

\indent\indent \emph{Proof.}  Please see Appendix \ref{section:proposition48dsdfa}.

Proposition \ref{proposition:decrease} implies that, under \emph{\textbf{Case 2}} and \emph{\textbf{Case 3}}, only if there exists an unique perturbation ratio and the target-perturbation branch is perturbed to that value, the dimension of the stealthy attack space $dim(\mathcal{A}_s)$ increases 1. Based on this result, if $\Delta \gamma = 0$ when we perturb a target-perturbation branch to a certain value, then this result almost remains the same regardless of the perturbation magnitude. Further, we obtain a result about the increase of the security factor.

\begin{theorem}\label{theorem:increasetheorem}
\emph{Under all cases, }$ \Delta \gamma = 1$ \emph{for any} $\lambda >0 $ ($\lambda \neq 1$)\emph{ if and only if there exists} $\lambda = \lambda^*$ ($\lambda^* \neq 1$)\emph{ such that }$\Delta \gamma = 1$.
\end{theorem}

\indent\indent \emph{Proof.} Please see Appendix \ref{section:theorem49bbuda}.

Theorem \ref{theorem:increasetheorem} implies that if the security factor $\gamma$ increases 1 when we perturb one more branch, then this result will not be changed with the variation of the susceptance perturbation magnitude. In other words, once we find that the dimension of the stealthy attack space $dim(\mathcal{A}_s)$ decreases 1 in a trial, then this result will not change in the other trials. Therefore, we can determine the security factor $\gamma$  using only one tested perturbation ratio when $\Delta \gamma = 1$.

In summary, the susceptance perturbation magnitude almost does not affect the change of the dimension of the stealthy attack space $dim(\mathcal{A}_s)$ when one more branch is perturbed. Thus, we can determine the dimension of the stealthy attack space after we have perturbed the branches to certain ratios, without worrying about the impact of the perturbation magnitude, especially the case when $\Delta\gamma = 1$. Note that this is very useful for the selection of the target-perturbation branches. We present details about this issue in the next subsection.

\subsection{Guidance on the construction of an effective MTD}\label{MTDconstruction}
In this section, we first consider the optimal selection of the set of target-perturbation branches. Then, we discuss the increasing generation cost caused by MTD.

\subsubsection{Target-perturbation branch selection}\label{section:branchselection}
Here we propose an algorithm for maximizing the security factor $\gamma$ (i.e., minimizing the dimension of the stealthy attack space $\mathcal{A}_s$) and covering the largest number of buses, with a given number of D-FACTS devices.  We briefly introduce the algorithm in the following.
\begin{algorithm}[htbp]
 \caption{Minimizing the dimension of the stealthy attack space}\label{maximumsecurtiyfactor}
 \KwData{The number of D-FACTS devices $n_d$; The set of branches that can be perturbed $\mathcal{L}_p$}
 Initialization: randomly rearrange $\mathcal{L}_p$ as $\mathcal{L}_p = \{k'_1, k'_2, \cdots, k'_p\}$, $r =0$, $r' = 0$\;
 \For{each $k'_t \in \mathcal{L}_p$}{
  Perturb $k'_t$ to a random ratio $\lambda$\;
  \If{$r = n_d$}{
   break \;
   }
   Compute $\Delta \gamma = R([\mathbf{H}~\mathbf{H}'']) - R([\mathbf{H}~\mathbf{H}'])$ \;
  \If{$\Delta \gamma = 1$}{
   put $k'_t$ in the branch set $\mathcal{L}^1_d$\;
   $r = r + 1$\;
   }
 }
 \eIf{$r < n_d$}{
   Enter Algorithm \ref{coverage}\;
   $\mathcal{L}_d = \mathcal{L}^1_d \cup \mathcal{L}^2_d $ \;
 }{$\mathcal{L}_d = \mathcal{L}^1_d$ \;
 }
 \KwResult{The set of target-perturbation branches $\mathcal{L}_d$}
\end{algorithm}

For the sake of power transfer quantity and quality, the branch parameters of some branches cannot be perturbed. Therefore, the set of candidate branches input to Algorithm 1 may not be the set of all branches in the power network. Thus, we only need to seek for an optimal deployment strategy in the set of branches that can be perturbed. With the loop (form line 2 to line 12) in Algorithm 1, we traverse all perturbable branches until we obtain the maximum $\gamma$. This process computes a set of target-perturbation branches that can minimize the stealthy attack space. In the loop, we determine whether the branch should be selected or not from line 8 to line 11. Once $\Delta \gamma =1$, the branch is selected as a target-perturbation branch; otherwise, it is not. When the iterations in Algorithm 1 have finished, we determine whether we should go into Algorithm 2 with the condition given in line 13. That is, if $\gamma$ is maximized and all the D-FACTS devices have been used, then the algorithm is closed and the output $\mathcal{L}^1_d$ is the set of target-perturbation branches. If $\gamma$ is maximized but there still exist unused D-FACTS devices, the algorithm enters Algorithm 2 for maximizing the covered buses.

Algorithm 2 starts with the rest candidate branches except those have been selected for maximizing $\gamma$. It incrementally searches for branches that cover new buses. The target-perturbation branch is selected according to the following process. First of all, we select the branches that cover two new buses (line 7 to line 10). Then, if there still exist unused D-FACTS devices, we select the branches that cover one new bus (line 16 and line 17). At last, if there still exist unused D-FACTS devices, we select the branches from the rest candidate branches (line 19 to line 21).

\begin{algorithm}[h!]
 \caption{Covering the largest number of buses}\label{coverage}
 Compute: $\mathcal{L}'_p = \mathcal{L}_p \setminus \mathcal{L}^1_d$; the set of buses $\mathcal{I}$ covered by branches in $\mathcal{L}^1_d$\;
 \For{each $k''_t=\{i''_t, j''_t\} \in \mathcal{L}'_p $}{
  \If{$r + r' = n_d$}{
   break \;
   }
  \If{$i''_t$ or $j''_t$ $\notin \mathcal{I}$}{
   \eIf{$i''_t$ and $j''_t$ $\notin \mathcal{I}$}
   {put $k''_t$ in the branch set $\mathcal{L}^2_d$ \;
    $r' = r' +1$ \;
   }{put $k''_t$ in the branch set $\mathcal{L}^3_d$ \;}
   }
 }
 \If{$r + r' < n_d$}{
 \eIf{$n_d- r - r' \leq \mid \mathcal{L}^3_d \mid$ }{
 Select $n_d- r - r'$ branches from $\mathcal{L}^3_d$ and put them into the branch set $\mathcal{L}^2_d$\;}{ Put all branches in $\mathcal{L}^3_d$ into $\mathcal{L}^2_d$\;
 Select $n_d- r - r' - \mid\mathcal{L}^3_d \mid$ branches from $\mathcal{L}'_p \setminus \mathcal{L}^2_d$ and put them into $\mathcal{L}^2_d$\;}}
 Return $ \mathcal{L}^2_d$ \;
\end{algorithm}

The computation time requirement for the rank operation of $\Delta\gamma=R([\mathbf{H}~\mathbf{H}''])-R([\mathbf{H}~\mathbf{H}'])$ is $\mathcal{O}(mn^2)$, where $m$ is the number of measurements and $n$ is the number of buses in the power network. The runtime for Algorithm 2 is $\mathcal{O}(l)$, where $l$ is the number of branches in the power network. Considering the worst case that the rank operation would be executed $l$ times in the loop, the time complexity for Algorithm 1 is $\mathcal{O}(lmn^2)$. Note that Algorithm 2 is not executed if the condition given in line 13 of Algorithm 1 is not satisfied. In fact, the above algorithm only outputs an alternative set of target-perturbation branches. There may be a lot of candidates that are also satisfied (see Section \ref{section:selectionbranchdnosanda}). Therefore, we can dynamically change the perturbed branches and maintain the effectiveness of MTD. On the other hand, we can realize MTD on the basis of already deployed D-FACTS devices for saving the additional infrastructure cost.

\subsubsection{Reducing the operation cost}\label{section:decreasecost}
After determining the set of branches that should be perturbed, a following problem that we should consider is to reduce the operation cost by appropriately setting their perturbation magnitudes. Optimal power flow (OPF) seeks to optimize the operation of an electric power system subject to the physical constraints imposed by electrical laws and engineering limits \cite{frank2016opf}. It outputs the minimum generation cost for the given loads by adjusting the power flows. This generation cost can represent the operation cost caused by the system changes of MTD. Here, OPF is stated as follows:
\begin{equation}\label{opf}
\begin{aligned}
  &\min_{p^g_i, i \in \mathcal{G}}~~~~\sum_{i \in \mathcal{G}}C_i(p^g_i) ~~~~~~~~~~~~~~~~\\
   &s.t.~~~~~~~p^g_i - p^l_i = \mathbf{B}\bm{\theta}, ~ i \in \mathcal{N} ~~~~~~~\\
  &~~~~~~~~~~~p^{min}_i \leq p^g_i  \leq p^{max}_i, ~i \in \mathcal{G} \\
  &~~~~~~~~~~f^{min}_{ij} \leq f_{ij} \leq f^{max}_{ij}, ~\{i,j\} \in \mathcal{L}\\
  &~~~~~~~~~~\lambda^{min}_{ij} \leq \lambda_{ij} \leq \lambda^{max}_{ij}, \{i, j\} ~ \in \mathcal{L} \\
\end{aligned}
\end{equation} where $C_i(p^g_i)$ is the cost function, $\mathcal{G}$ is the set of generators, $p^g_i$ is the real output power, $p^l_i$ is the load, $f_{ij}$ is the branch active power flow, $\lambda_{ij}$ is the susceptance perturbation ratio. In the above optimization problem, the decision variable is the output generation level of each generator, and the cost function is a quadratic function. The first constraint is about the nodal power balance constraint, i.e., the power injections must be equal to the power consumptions. The second and third constraints are about the limits about the output generation and branch active power flow, respectively. The fourth constraint is newly added here for the limits of the susceptance perturbation. We can see that the matrix $\mathbf{B}$ contained in the first constraint contains the branch perturbation parameters. Therefore, this optimization problem is correlated with the branch perturbation magnitude. Since the objective function is convex and the constraints are differentiable, the OPF problem is a typically convex optimization problem \cite{wood1996system} \cite{boyd2004convexoptimization}, which can be solved by the non-linear programming solver \emph{fmincon} in MATLAB.

\section{Simulation Results}\label{section:simulation}
\begin{figure}[h]
\begin{center}
\includegraphics[width=0.22\textwidth]{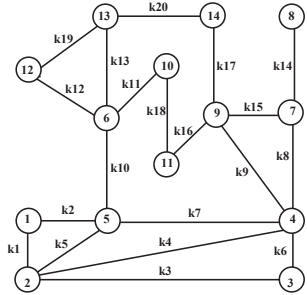}
\caption{The IEEE 14-bus power system.}\label{fig:14bus_pfi}
\end{center}
\vspace{-0.6cm}
\end{figure}

In this section, we evaluate our findings about MTD using an illustrative IEEE 14-bus power system (Fig. \ref{fig:14bus_pfi}) and the IEEE 30-bus, 57-bus, 118-bus and 145-bus power systems. All simulations are based on the fully measured power system and carried out in MATLAB. We assume that all meters are subject to the same noise 
distribution, namely the Normal distribution $\mathcal{N}(0, \sigma^2)$, if the measurement noises are considered. The threshold $\tau$ (see Section \ref{section:systemmodel}) is set as $\sigma \sqrt{\chi^2_{m-n, \alpha}}$ \cite{wood1996system}, where $m - n$ is the freedom degree of the Chi-square distribution and $\alpha$ is the false alarm rate (which is 0.05). In practice, each branch susceptance must be within given limits, namely $b^{min}_{ij} \leq b_{ij} \leq b^{max}_{ij}$ \cite{tian2018hidden}. The authors in \cite{morrow2012perturbation1} have proved that if the perturbations are within 20\% of the impedance, then there are sufficiently large number of perturbation cases that can restrict power losses within 1\%. Thus, it is feasible to perturb the branch susceptances within 20\% maximum change. In this paper, for the perturbation ratio $\lambda$, it is constrained to be within $[0.8, 1.2]$.

\subsection{Effectiveness of MTD}\label{section:completenessanalysis}

\subsubsection{The dimension of the stealthy attack space vs. branch perturbations}\label{section:changeofspace}
\begin{figure*}[htbp]
\begin{minipage}[t]{0.32\linewidth}
\centering
  \includegraphics[width=6cm]{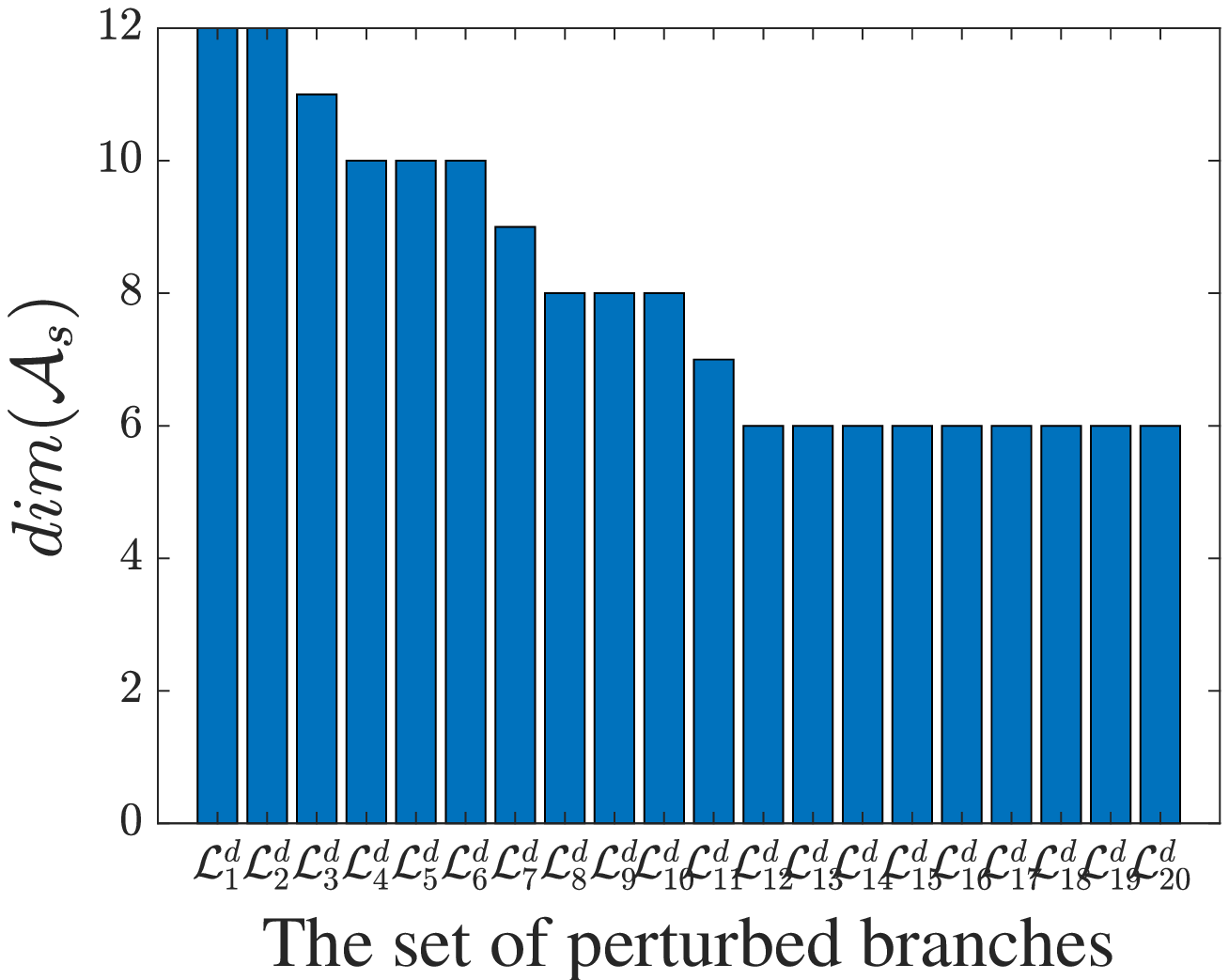}
 \caption{The dimension change of the stealthy attack space $\mathcal{A}_s$ with respect to the perturbed branches.}\label{fig:rank_plot}
\end{minipage}%
\hfill
\hfill
\begin{minipage}[t]{0.32\linewidth}
\centering
 \includegraphics[width=6cm]{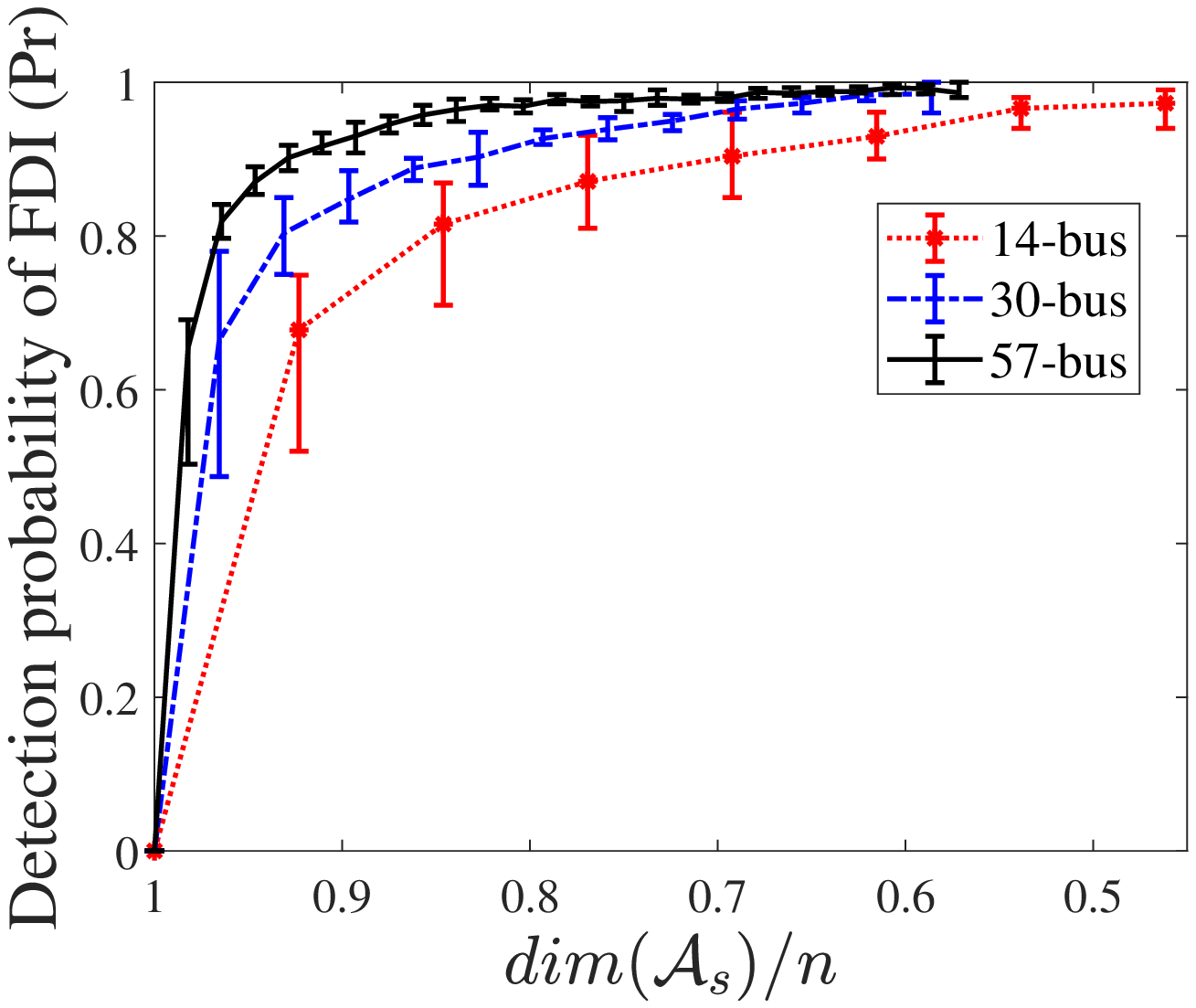}
 \caption{The detection probability of FDI attacks with respect to the dimension of the stealthy attack space.}\label{fig:detection_probability_multiple_systems}
 \end{minipage}%
\hfill
\hfill
\begin{minipage}[t]{0.32\linewidth}
\centering
  \includegraphics[width=6cm]{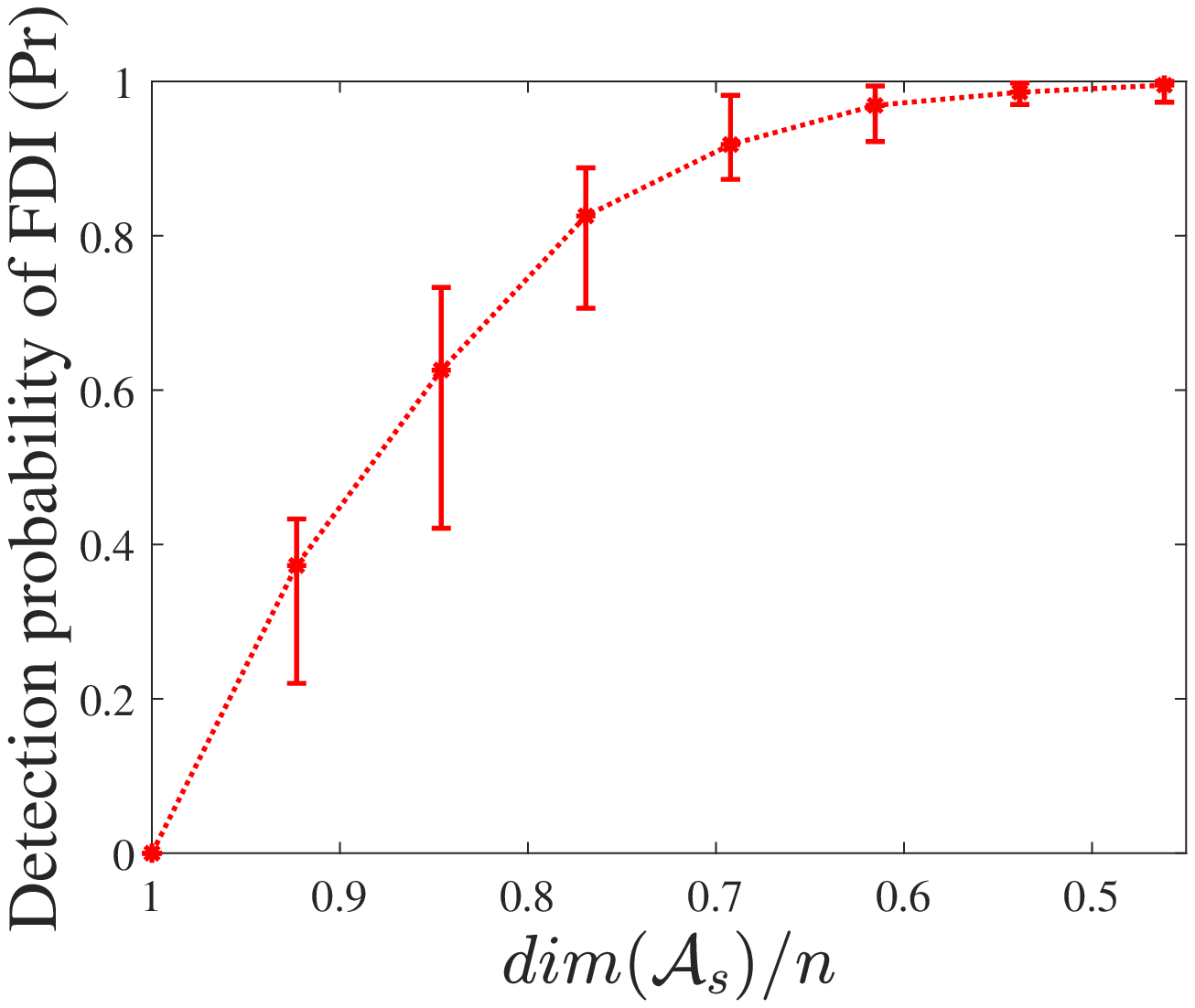}
  \caption{The detection probability of FDI attacks when only 3 state variables can be modified by the attacker.}\label{fig:detection_probability_with_fixed_state}
\end{minipage}%
\vspace{-0.2cm}
\end{figure*}

Taking the IEEE 14-bus power system (Fig. \ref{fig:14bus_pfi}) as an example, we analyze the dimension change of the stealthy attack space ($\mathcal{A}_s$) when we increase the number of perturbed branches. We successively perturb the set of branches as $\mathcal{L}^d_1 =\{k_1\}$, $\mathcal{L}^d_2 =\{k_1, k_2\}$, $\mathcal{L}^d_3 =\{k_1, k_2, k_3\}$, $\cdots$. Fig. \ref{fig:rank_plot} shows the simulation results. We can see that the dimension of the stealthy attack space decreases when more branches are perturbed. But $dim(\mathcal{A}_s)$ cannot reach 0, because the 14-bus power system only has 20 branches, which is less than $2n = 26$. Even though we perturb all branches, the dimension of the stealthy attack space is 6, which is the smallest stealthy attack space we can achieve in this 14-bus power system. Consistent with the equation (\ref{dim}), the smallest dimension of the stealthy attack space is equal to $2n-l = 2*13-20 = 6$.

\begin{figure*}[htbp]
\begin{minipage}[t]{0.32\linewidth}
\centering
  \includegraphics[width=6cm]{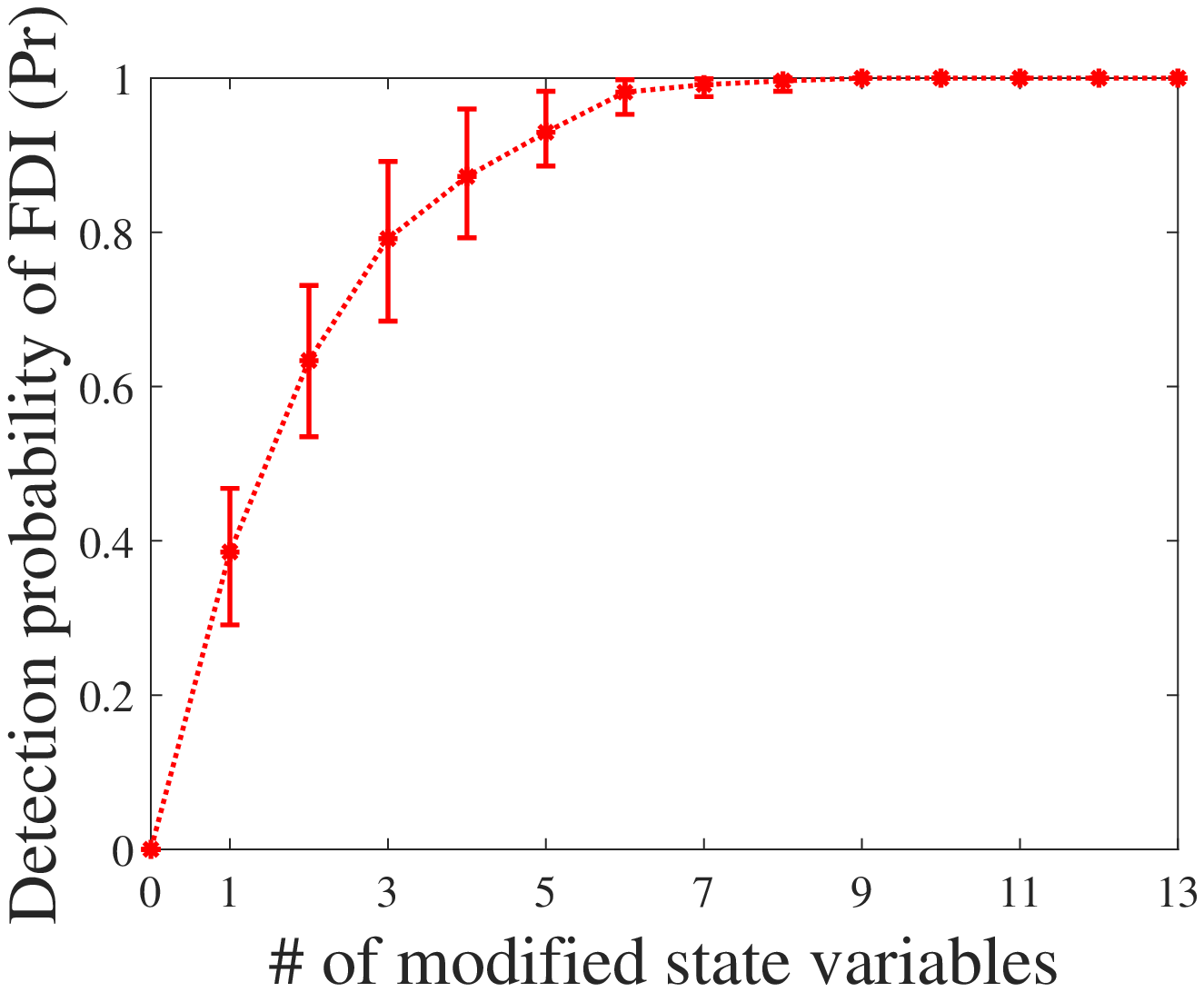}
 \caption{The detection probability of FDI attacks when the dimension of the stealthy attack space is fixed.}\label{fig:detection_probability_fixed_space}
\end{minipage}%
\hfill
\hfill
\begin{minipage}[t]{0.32\linewidth}
\centering
  \includegraphics[width=6cm]{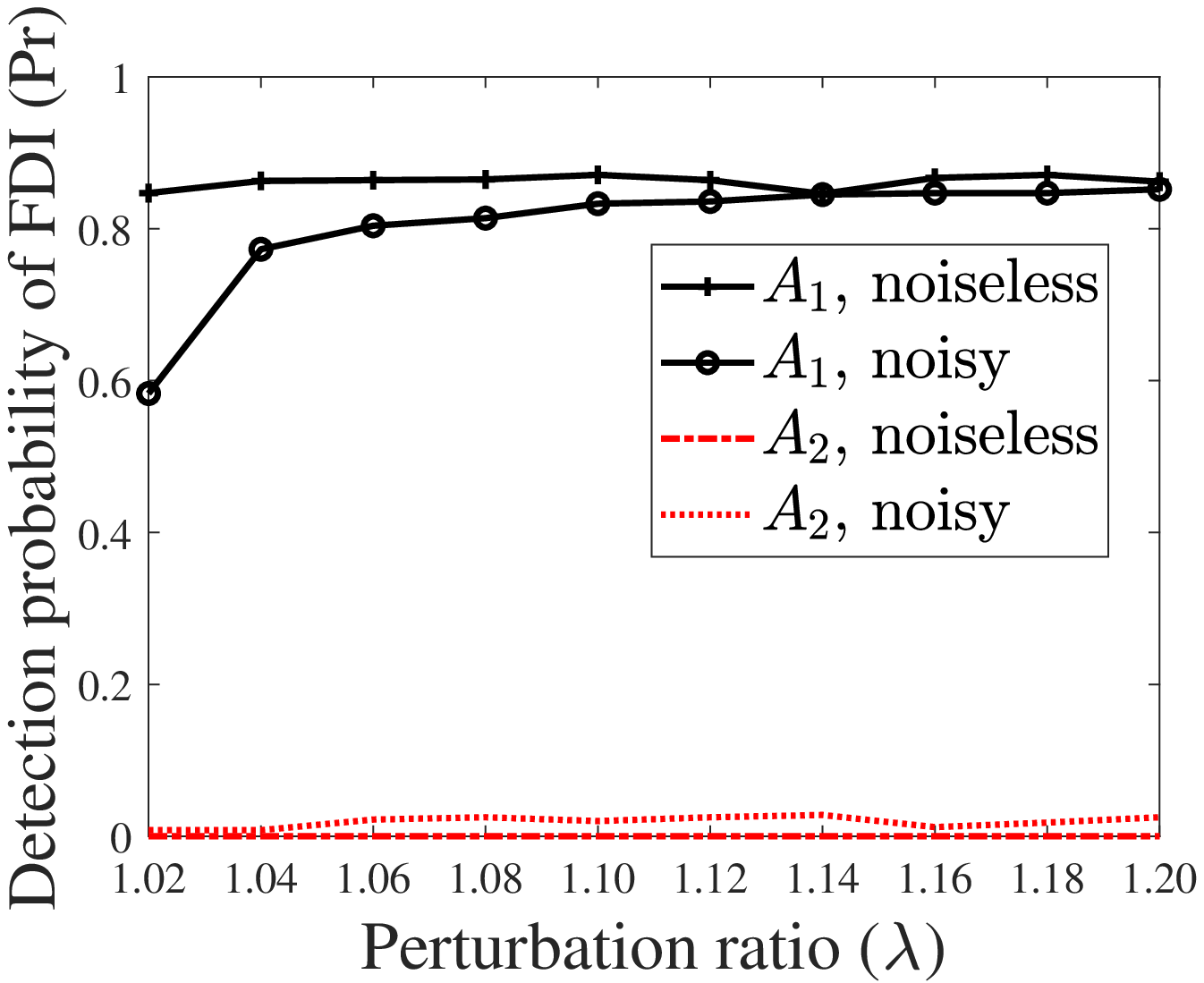}
 \caption{The detection probability of FDI attacks with $ c_i \neq 0$ for $ i \in \mathcal{M}_{mtd}$ ($A_1$) and $c_i = 0$ for all $ i \in \mathcal{M}_{mtd}$ ($A_2$).}\label{fig:Detection_probability_notcarecover}
 \end{minipage}%
\hfill
\hfill
\begin{minipage}[t]{0.32\linewidth}
\centering
  \includegraphics[width=6cm]{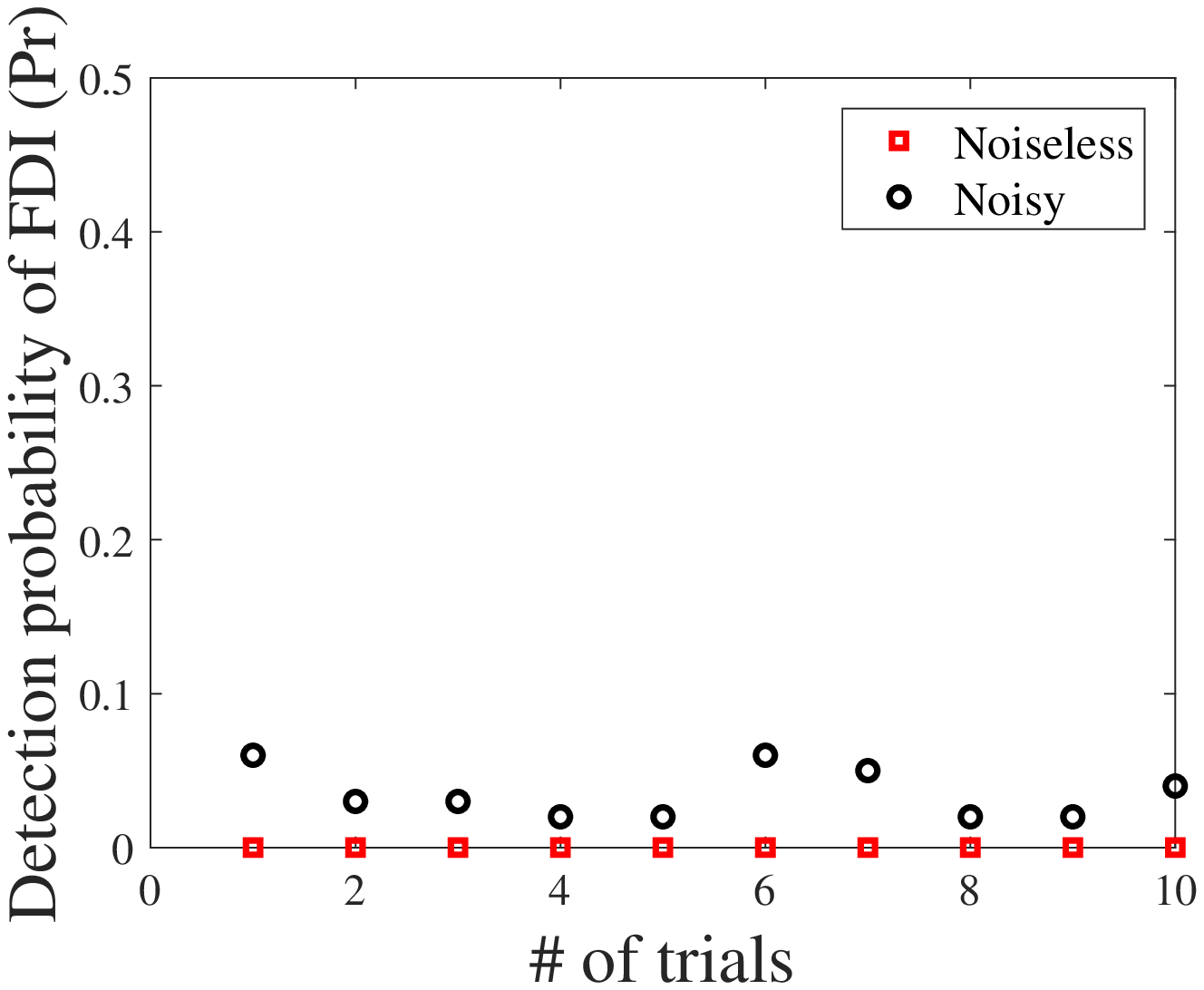}
 \caption{The detection probability of FDI attacks vs. different sets of perturbed branches with only bus 8 is attacked.}\label{fig:detection_probability_bus8}
\end{minipage}%
\vspace{-0.4cm}
\end{figure*}


Moreover, we perturb all branches in the IEEE 30-bus, 57-bus, 118-bus and 145-bus power systems. Table \ref{table:resultsafterall} shows the smallest dimension of the stealthy attack space we obtain. We also give the number of branches for each power system in this table. We can see that all the smallest dimensions of the stealthy attack space are not zero, which indicates that none of the power systems support constructing complete MTD. Even though the IEEE 145-bus power system has 453 branches, which is larger than $2n = 288$, the dimension of the stealthy attack space is 10 after perturbing all its branches. In our opinion, one reason for this result is that the IEEE 145-bus power system contains 7 buses that are only connected by a single branch. The simulation result indicates that is difficult to meet the conditions required for achieving a complete MTD in practice.

\begin{table}[htbp]
\centering
 \caption{The dimension of $\mathcal{A}_s$ after perturbing all branches}\label{table:resultsafterall}
 \begin{tabular}{c|cccc}
    \hline
 IEEE test system & 30-bus &  57-bus & 118-bus & 145-bus\\
   \hline
  $\#$ branches & 41 & 80 & 186 & 453   \\
     \hline
$dim(\mathcal{A}_s)$& 18 & 31 & 40 & 10 \\
    \hline
 \end{tabular}
\end{table}

\subsubsection{MTD's effectiveness for thwarting FDI attacks}\label{section:effectivenessofmtd}
Next, we exploit MTD's effectiveness for thwarting FDI attacks with respect to the dimension of the stealthy attack space. The attack vector is constructed with the form of $\textbf{a} = \mathbf{H}\textbf{c}$, where $\mathbf{H}$ is a measurement matrix before MTD. We sample the value of the element $c_i$ in $\textbf{c}$ from a uniform distribution $\mathcal{U}(-dm, dm)$, where $dm$ is the maximum magnitude of the injected bias into the state variable. Here $dm$ is 0.1. The modified state variables are uniformly selected from the bus set, i.e., the non-zero elements in $\textbf{c}$ are uniformly selected. We assume that all branches in the power system can be perturbed, and the measurements are noiseless. The perturbation ratio is randomly chosen within $[0.8, 1.2]$. For each setting, we repeat random attacks for 1000 times based on Monte Carlo simulations, and estimate the detection probability of the FDI attack constructed with $\mathbf{H}$ as
\begin{equation}\label{detectionprobability}
  \text{Pr} =\frac{\#~\text{of detected trials}}{1000},
\end{equation}
where \emph{\# of detected trials} means the number of FDI attack vectors detected by the BDD after MTD. For consistency, we use the metric $\frac{dim(\mathcal{A}_s)}{n}$ to measure the change of the dimension of the stealthy attack space. Note that for a fixed dimension of the stealthy attack space, there are several branch-perturbation schemes for realizing MTD.

We use the IEEE 14-bus, 30-bus and 57-bus power systems with default settings and data in MATPOWER. Fig. \ref{fig:detection_probability_multiple_systems} shows the simulation results. We can see that if the dimension of the stealthy attack space is smaller, the detection probability of FDI attacks is larger. For example, in the 14-bus power system, the detection probability of FDI is more than 90\% when $\frac{dim(\mathcal{A}_s)}{n}$ is reduced to 0.6. Moreover, we find that if the system size is larger, the curve is more smooth and the detection probability of FDI attacks is larger given the same $\frac{dim(\mathcal{A}_s)}{n}$. The simulation results highlight that the smaller the dimension of the stealthy attack space is, the better performance the MTD achieves in terms of thwarting FDI attacks.

Besides, we consider the case when the attacker has limited resources to modify state variables. We take the IEEE 14-bus power system as an example. First, we assume that the attacker can only modify 3 state variables, that is, there are only 3 non-zero elements in $\textbf{c}$. We analyze the change of the detection probability of FDI attacks with respect to the dimension of the stealthy attack space. For each setting, we repeat random attacks for 1000 times based on Monte Carlo simulations. That is, for each time, the attacked state variables and the attack vector are randomly generated. Fig. \ref{fig:detection_probability_with_fixed_state} shows the simulation result. We can see that the detection probability increases with $\frac{dim(\mathcal{A}_s)}{n}$. Again, it proves that a smaller stealthy attacks space is more effective to thwart FDI attacks. Second, we assume that the dimension of the stealthy attack space is fixed as 10, while the number of modified state variables is varied. Note that there are several combinations for a specific number of state variables. For each setting, we repeat the simulation for 1000 times and uniformly select a set of state variables for each time. The simulation result is plotted in Fig. \ref{fig:detection_probability_fixed_space}. We can see that the detection probability of FDI attacks increases with the number of modified state variables. This indicates that the attacker is more ambitious, the MTD is more effective.

Moreover, we investigate the impact of FDI attacks constructed with the former measurement matrix on the buses that are not covered by the perturbed branches. We select $k_1$, $k_3$, $k_4$ and $k_7$ as the target-perturbation branches. They are perturbed to the same ratio during simulations. The perturbation ratios are set within $[1.02, 1.2]$ spaced by 0.02. The attack vector $\textbf{a} = \mathbf{H}\textbf{c}$ is constructed by considering two cases: $ c_i \neq 0$ for $ i \in \mathcal{M}_{mtd}$ and $c_i = 0$ for all $ i \in \mathcal{M}_{mtd}$, where $\mathcal{M}_{mtd}$ is a set of buses that are covered by the perturbed branches. We denote these two cases as ``$A_1$" and ``$A_2$," respectively. The measurement noise is fixed as $\sigma^2 =0.01$. For each setting, we repeat the simulation for 1000 times. For each time, we uniformly selected the non-zero elements in $\textbf{c}$. The simulation results are given in Fig. \ref{fig:Detection_probability_notcarecover}. We can see that, for case $A_1$, the detection probability of the FDI attack is more than 84\% when the measurements are noiseless. We believe that this large detection probabilities is because $\textbf{a} \notin S(\mathbf{H}')$ holds for most cases. When the measurements are noisy in case $A_1$, the detection probability of FDI attacks is disturbed by the noise. But it is still more than 60\% and increases with the perturbation ratio. For case $A_2$, we can see that, whether the measurements are noisy or not, the detection probability of FDI attacks is almost negligible. This indicates that the buses that are not covered by the perturbed branches are vulnerable to FDI attacks. Therefore, to improve the effectiveness of MTD, the perturbed branches should cover as many buses as possible.
\begin{figure}[h]
\begin{center}
  \includegraphics[width=6cm]{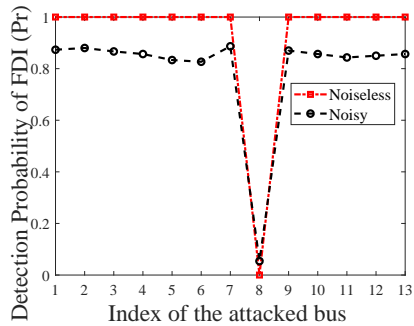}
 \caption{The attacks on bus 8 can always bypass the BDD after MTD in the IEEE 14-bus power system.}\label{fig:singlebranch_bus}
\end{center}
\vspace{-0.4cm}
\end{figure}
\subsubsection{Impact of the bus that is only connected by a single branch}\label{section:impactofsinglebus}
Next, we consider the attack on the bus that is only connected by a single branch. Here we validate our finding using the IEEE 14-bus power system. We find that bus 8 is only connected by branch $\{7,8\}$. First, we only inject errors into the state variable of bus 8. We test this attack for 10 trials. For each trial, we select 1000 sets of perturbed branches except for branch $\{7,8\}$ randomly. The noise variance is fixed as 0.01. Fig. \ref{fig:detection_probability_bus8} shows the simulation results. We can see that the attacks on bus 8 can always bypass the BDD after MTD when the measurements are noiseless. And for the noisy case, the detection probability of this attack is almost negligible (around 0.05). By contrast, we conduct another 13 trials but focus on attacking different buses. For the $i$th trial, we only modify the state variable of bus $i$. In these trials, we perturb all branch susceptances. Fig. \ref{fig:singlebranch_bus} shows the simulation results. We can see that, no matter the measurements are noiseless or noisy, we have a negligible detection probability of the FDI attack when the state variable of bus 8 is modified. But it can be detected when we attack the other buses. The above results highlight the weakness of the bus that is only connected by a single branch and prove that we can never achieve a complete MTD when the power network contains such a bus.

Next, we consider the induced generation cost when attacking the bus that is only connected by a single branch. Considering the IEEE 30-bus power system, there are 3 buses that are only connected by a single branch. And one of them is a load bus (bus 26). In the following, we exploit the increasing generation cost when the attacker compromises this load bus. The objective function is linear with the value of generation, i.e., $C_i(p^g_i) = \mu_i p^g_i$ [see the OPF problem (\ref{opf})]. The parameters about the generators are shown in Table \ref{table:opfgenerationcostnwe}. And all active power flows are limited to 500 MW. In this case, we perturb all branches with the perturbation ratios sampled within $[0.8, 1.2]$. The default setting and load data provided in MATPOWER are used. First, we use \emph{fmincon} in MATLAB to compute an optimal generation dispatch result. Then, the measurements associated with bus 26 are corrupted. It results in deceiving the amount of the load on bus 26 transmitted to the control center for solving the OPF problem. The simulation results are shown in Table \ref{table:opfgenerationcostresults}. $L_0$ is the original load at bus 26. We can see the generation cost increases with the corrupted load. Actually, the load attack may also lead the system to a non-optimal generation dispatch, and the worst, may cause load shedding \cite{liang2016vullr}.

\begin{table}[htbp]
\centering
 \caption{Parameters of the generators}\label{table:opfgenerationcostnwe}
 \begin{tabular}{c|cccccc}
 \hline
 Generation bus & 1  & 2 & 13 & 22 & 23 & 27\\
  \hline
 $P^g_{max}$ & 100  & 100 & 100 & 100 & 100 &100\\
 \hline
 $\mu_i$(\$/MWh)& 20 & 30 & 20 & 20 & 30 & 20\\
 \hline
 \end{tabular}
\end{table}

\begin{table}[htbp]
\centering
 \caption{Increasing in generation cost when the load bus that is only connected by a single branch is attacked }\label{table:opfgenerationcostresults}
  \scalebox{0.85}{
 \begin{tabular}{c|cccccc}
  \hline
 Communicated load  & $L_0$  & 1.2$L_0$ & 1.4$L_0$ & 1.6$L_0$ & 1.8$L_0$ & 2.0$L_0$\\
  \hline
 Generation cost ($10^{3}\$/h$) & 3.784  & 3.798 & 3.812 & 3.826 & 3.840 &3.854\\
 \hline
  Increasing rate & $0$  & 0.37\% &0.74\% & 1.11\% & 1.48\% & 1.85\%\\
 \hline
 \end{tabular}}
\end{table}

\subsection{Impact of the susceptance perturbation magnitude on the dimension of the stealthy attack space}\label{section:impactoffdimag}
To demonstrate the results given in Section \ref{section:reduceattackspace}, we analyze the dimension change of the stealthy attack space by varying the perturbation ratio. We give six examples for the three cases stated in Section \ref{section:reduceattackspace} with the IEEE 14-bus power system. For each example, we randomly choose 10000 perturbation ratios within $[0.8, 1.2]$. At the beginning, we choose an initial set of perturbed branches. Then, we perturb one more branch $k_d$. Let $dim(\mathcal{A}_s)$ and $dim^*(\mathcal{A}_s)$ denote the dimension of the stealthy attack space before and after this perturbation, respectively. During simulations, we count the number of exceptions when $dim^*(\mathcal{A}_s)$ is changed. The simulation results are shown in Table \ref{table:resultsrank}. We can see that if the dimension of the stealthy attack space is reduced by 1, then this result always holds for all perturbation ratios in these three cases. If the dimension of the stealthy attack space remains the same after the perturbation of branch $k_d$, we do not find exceptions in the first (\emph{\textbf{case 1}}) and third (\textbf{\emph{case 2}}) example, i.e., this result always holds regardless of the perturbation ratio. But in the fifth (\emph{\textbf{case 3}}) example, we find the dimension of the stealthy attack space increases 1 when we set the perturbation ratio to be 0.88. In other words, $dim(\mathcal{A}_s)= dim^*(\mathcal{A}_s)$ does not always hold in \emph{\textbf{case 3}}. Moreover, we find that this exception happens when the perturbation ratio of branch $k_6$ is the same with that of branch $k_3$. Therefore, in order to avoid the change of stealthy attack space's dimension with the perturbation magnitude, we should not perturb the branches to the same ratio.
\begin{table}[htbp]
\centering
 \caption{Impact of the perturbation magnitude on the dimension of the stealthy attack space}\label{table:resultsrank}
  \scalebox{0.9}{
 \begin{tabular}{c|c|c|c|c}
    \hline
    Cases  & Initial set & $k_d$  & $dim^*(\mathcal{A}_s) - dim(\mathcal{A}_s)$ & \# Exceptions\\
    \hline
     \multirow{2}*{\textbf{\emph{Case 1}}} & \{$k_1, k_2, k_3, k_4$\}   & $k_{14}$ &0 & 0  \\ \cline{2-5}
      ~ &\{$k_1, k_2, k_3, k_4$\}  &  $k_{16}$  & -1& 0 \\
      \hline
      \multirow{2}*{\textbf{\emph{Case 2}}} & \{$k_1, k_2, k_3, k_4$\} & $k_5$ &0 & 0  \\ \cline{2-5}
      ~ &\{$k_5, k_6, k_8$\}  &  $k_1$  & -1& 0 \\
      \hline
      \multirow{2}*{\textbf{\emph{Case 3}}} &  \{$k_1, k_2, k_3, k_4$\} & $k_6$  &0 & 1 \\ \cline{2-5}
      ~ &\{$k_1, k_6$\}  &  $k_4$  & -1& 0 \\
      \hline
 \end{tabular}}
\end{table}

\subsection{Guidance on constructing an effective MTD}
\subsubsection{Target-perturbation branch selection}\label{section:selectionbranchdnosanda}

\

\begin{figure}[!htbp]
  \centering
  \subfigure[]{
    \label{fig:output_from_our_algorithm} 
    \includegraphics[width=1.50in]{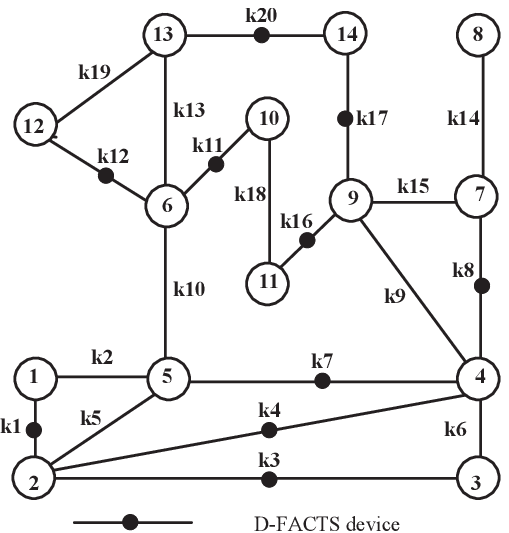}}
  \subfigure[]{
    \label{fig:not_chose_correctly} 
    \includegraphics[width=1.50in]{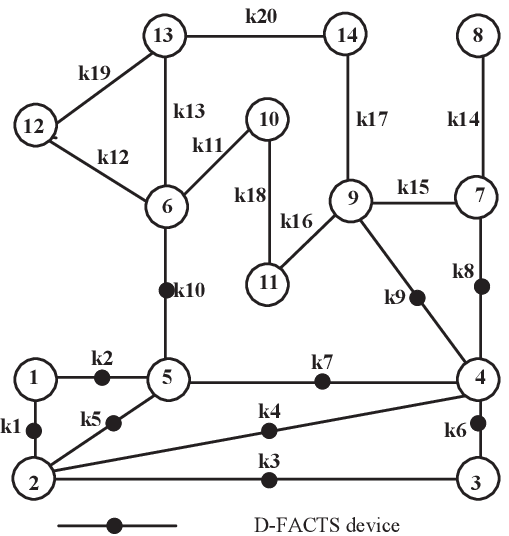}}
  \caption{Comparison of the optimal deployment of D-FACTS devices using our algorithm and the optionally selected target-perturbation branches. (a) Output from the Algorithms given in Section \ref{section:branchselection}; (b) Optionally selected target-perturbation branches. In the first case, dim$(\mathcal{A}_s) = 6$ and the number of covered buses is 13; while in the second case, dim$(\mathcal{A}_s) = 8$ and the number of covered buses is 8.}
  \label{fig:example} 
  \vspace{-0.2cm}
\end{figure}

\textbf{Evaluation of our algorithm.} First, with the IEEE 14-bus power system, we evaluate the selection of the set of target-perturbation branches with Algorithm 1 and Algorithm 2. A total number of 10 D-FACTS devices are given for constructing/realizing MTD. We assume that all branches in this power system can be perturbed. With the aim to minimize the dimension of the stealthy attack space and maximize the number of covered buses, we adopt Algorithm 1 and 2 in Section IV-D1. From Algorithm 1, we obtain a set of target-perturbation branches as $\{k_1, k_3, k_4, k_7, k_8, k_{11}, k_{12}\}$. From Algorithm 2, we obtain a set of target-perturbation branches as $\{k_{16}, k_{17}, k_{20}\}$. We get the final set of the target-perturbation branches by combining these two sets. With this MTD, the dimension of the stealthy attack space is 6 and the number of covered buses is 13. The target-perturbation branches deployed with D-FACTS devices are shown in Fig. \ref{fig:output_from_our_algorithm}. For comparison, if we optionally perturb the set of branches $\{k_1, k_2, k_3, k_4, k_5, k_{6}, k_{7}, k_8, k_9, k_{10}\}$ (Fig. \ref{fig:not_chose_correctly}), the dimension of stealthy attack space is 8 and the number of the covered buses is 8. This indicates that the output from our algorithm is better than the optional selections.

\begin{figure*}[htbp]
\begin{minipage}[t]{0.32\linewidth}
\centering
  \includegraphics[width=6cm]{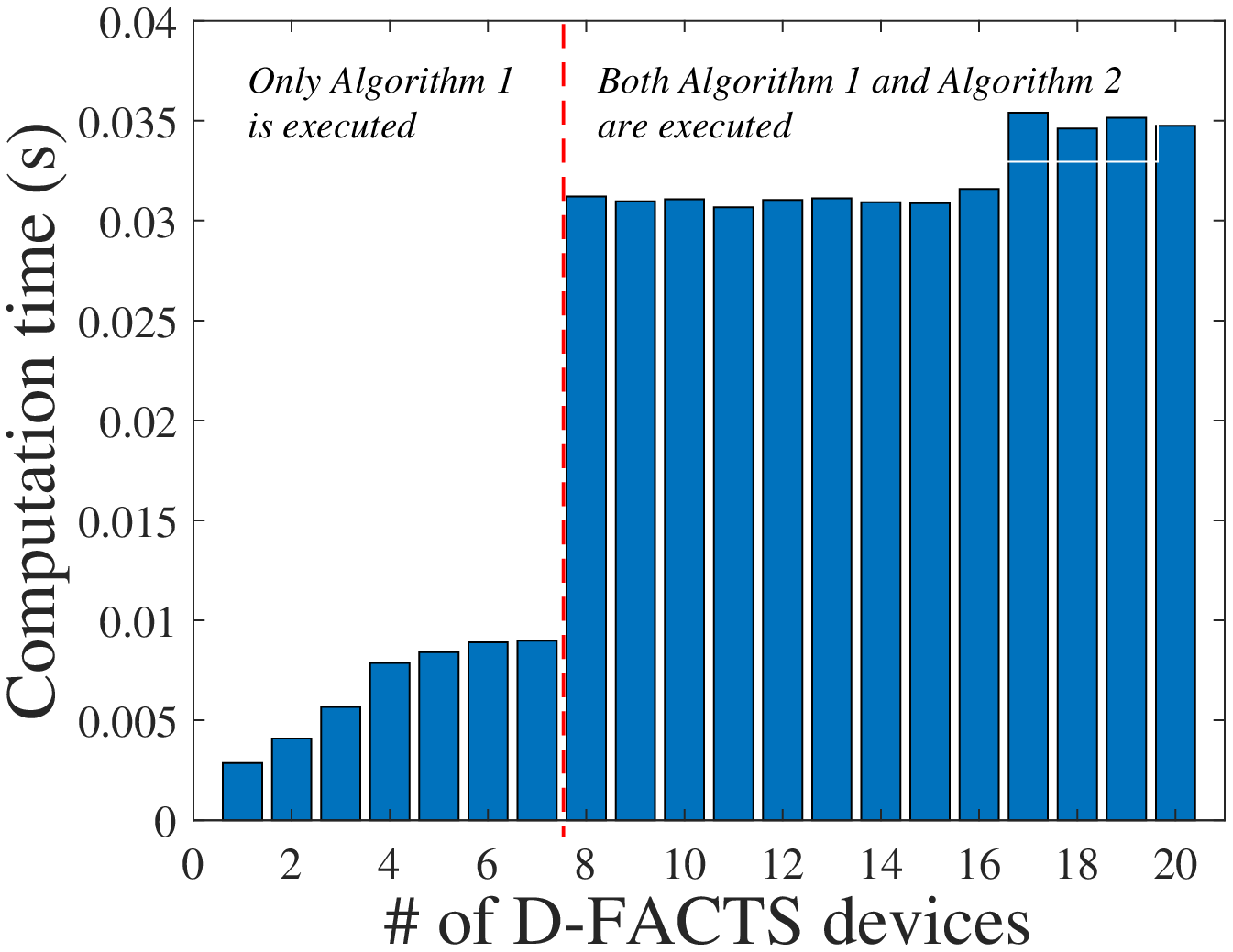}
 \caption{The variation of the computation time of the proposed algorithm with different numbers of D-FACTS devices (14-bus power system).}\label{fig:computation_time_14bus}
\end{minipage}%
\hfill
\hfill
\begin{minipage}[t]{0.32\linewidth}
\centering
 \includegraphics[width=6cm]{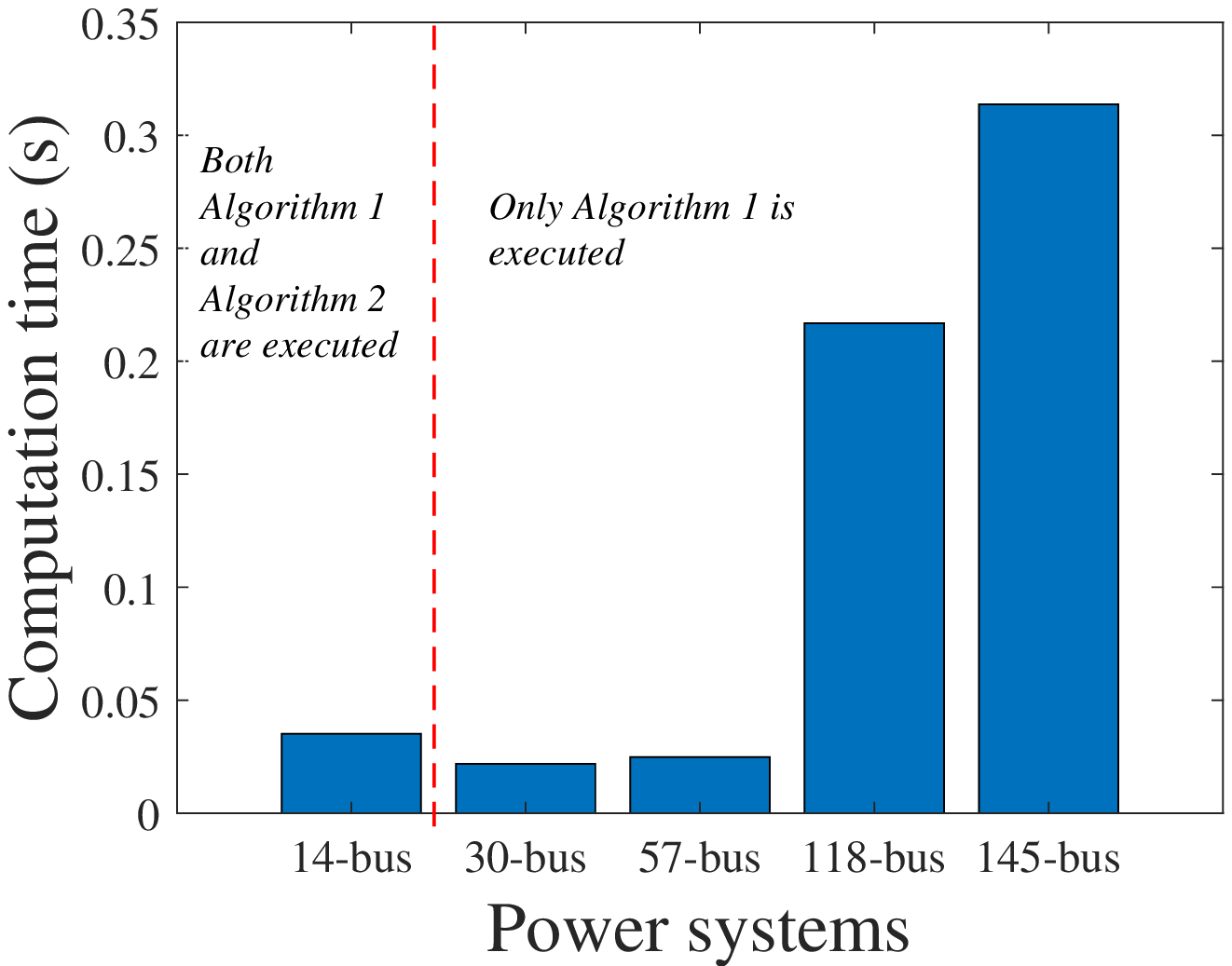}
 \caption{The variation of the computation time of the proposed algorithm with different power systems (given 10 D-FACTS devices).}\label{fig:computation_time_vsys}
 \end{minipage}%
\hfill
\hfill
\begin{minipage}[t]{0.32\linewidth}
\centering
 \includegraphics[width=6cm]{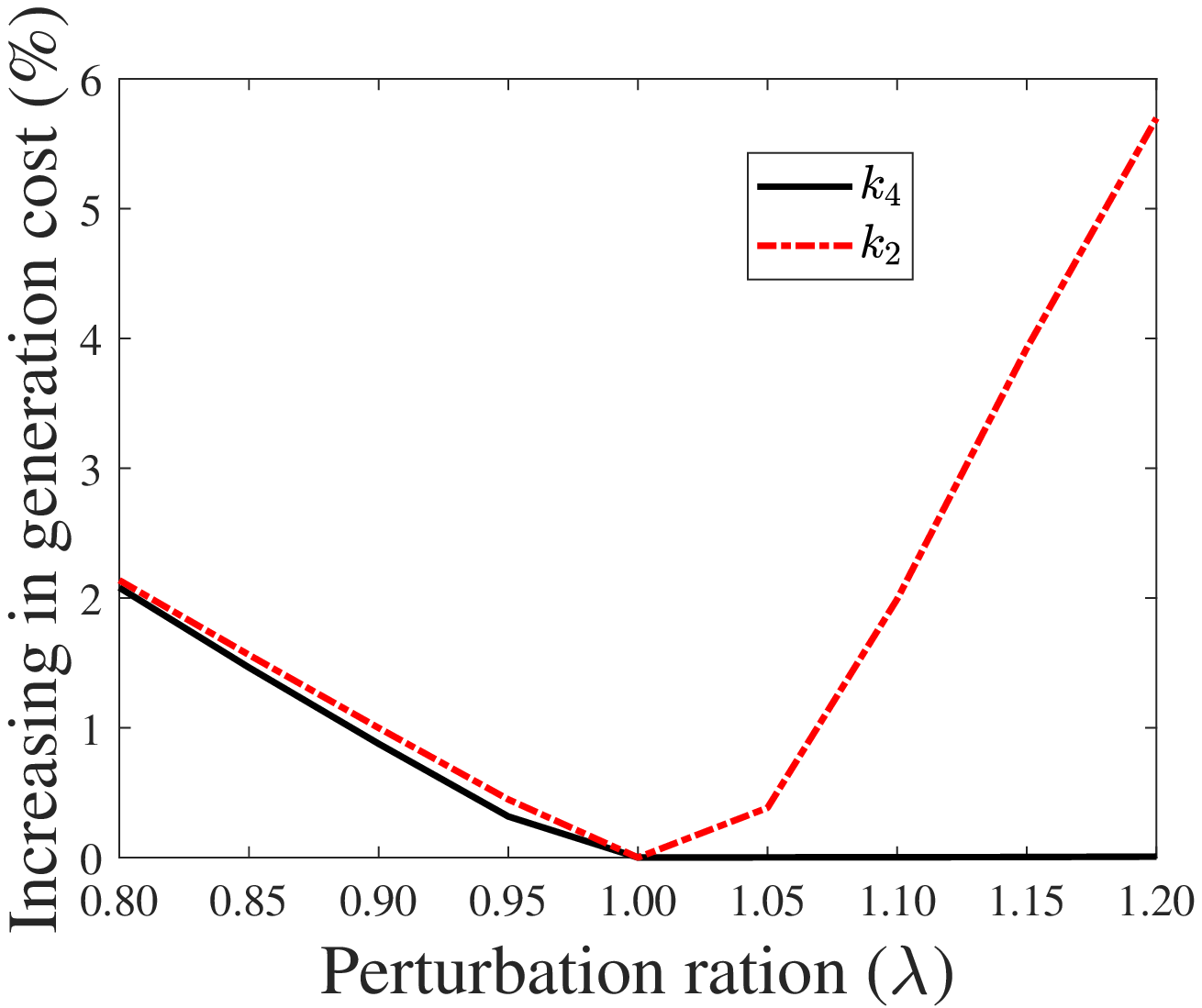}
 \caption{The increasing in generation cost by perturbing branch $k_2$ and $k_4$ to different ratios.}\label{fig:opf_cost_vary_bran}
\end{minipage}%
\vspace{-0.4cm}
\end{figure*}

In fact, if we start Algorithm 1 from different branches, we can obtain different sets of target-perturbation branches. We take the IEEE 14-bus power system as an example. Suppose that there are 9 D-FACTS devices for constructing MTD, and all branches in this 14-bus power system can be perturbed. We start Algorithm 1 from branch $k_2$ and $k_{16}$, respectively. Correspondingly, we obtain the sets of target-perturbation branches as $\mathcal{L}^{\text{from}~ k_2}_d = \{k_2, k_3, k_7, k_9, k_{10}, k_{11}, k_{15}, k_{16}, k_{19}\}$ and $\mathcal{L}^{\text{from}~ k_{16}}_d = \{k_2, k_4, k_6, k_8, k_{10}, k_{11}, k_{16}, k_{19}, k_{20}\}$, respectively. We can see that the sets of target-perturbation branches are different when we start Algorithm 1 from branch $k_2$ and $k_{16}$. But we find that the values of $\gamma$ and the number of covered buses are the same, i.e., $\gamma = 7$ and the number of covered buses is 13. This indicates that, even though the sets of target-perturbation branches output from our algorithm are different with different start points, they result in the same dimension of stealthy attack space and the same number of covered buses. Thus, we can dynamically change the perturbed branches and maintain the effectiveness of MTD. On the other hand, we can select the branches that are already deployed with D-FACTS devices for realizing MTD, which can help reducing the infrastructure cost.

\textbf{Computation time of the algorithm.} Moreover, we evaluate the computation time of the algorithm for the selection of target-perturbation branches. Note that Algorithm 2 is not executed if the condition given in line 13 of Algorithm 1 is not satisfied. We run the algorithm in a core i7 laptop, which has a 2.4GHz CPU and 8.0G memory. We assume that all branches can be perturbed in the adopted power systems. First, we use the IEEE 14-bus power system as an example. We vary the number of perturbed branches from 1 to 20. The computation time of our algorithm is shown in Fig. \ref{fig:computation_time_14bus}. We can see that the computation time is less than 10ms when the number of D-FACTS devices is less than 8, while it is more than 30ms when the number of D-FACTS devices is larger than 8. It seems that the computation time increases 3 times when the number of D-FACTS devices increases from 8 to 9. This is because only Algorithm 1 is executed when the number of D-FACTS devices is less than 8, i.e., the condition in line 13 of Algorithm 1 is not satisfied. While both Algorithm 1 and Algorithm 2 are executed when the number of D-FACTS devices is larger than 8. This indicates that sometimes it takes longer for executing Algorithm 2.

Second, we change the size of the power system while fix the number of D-FACTS devices as 10. Here we adopt the IEEE 14-bus, 30-bus, 57-bus, 118-bus and 145-bus power systems. The computation time of our algorithm is given in Fig. \ref{fig:computation_time_vsys}. We can see that the computation time increases from around 30ms (14-bus) to more than 300ms (145-bus), which indicates that the computation time of our algorithm increases with the system size. Moreover, we find that the computation time with the 14-bus power system is a little larger than that with the 30-bus power system. In our opinion, the reason is that, given 10 D-FACTS devices, both Algorithm 1 and Algorithm 2 are executed with the 14-bus power system, while only the Algorithm 1 is executed with the 30-bus, 57-bus, 118-bus and 145-bus power systems. Overall, we can efficiently compute the result (within 350ms in all given power systems).

\subsubsection{Reducing the operation cost}
Furthermore, we evaluate the impact of the perturbed branch and the susceptance perturbation magnitude on the increasing of the operation cost. For the OPF problem, we use the objective function as $C_i(p^g_i) = \mu_i p^g_i $, which is a linear generation cost model. With the IEEE 14-bus power system, the generators are installed at bus 1, 2, 3, 6 and 8,  and their parameters are shown in Table \ref{table:opfgenerationcost}. The active power flow limits of branch 1 is 160 MW and the other active power flows are limited to 60 MW. We assume that the optimal results are obtained at the beginning. Then, we analyze the increase of the generation cost when we perturb branch $k_2$ and $k_4$ to different ratios, respectively. The simulation results are plotted in Fig. \ref{fig:opf_cost_vary_bran}. We can see that the increasing of the generation cost under these two cases are different. When the branch susceptance is decreased, the increasing of generation cost by perturbing $k_4$ is slightly lower than that of by perturbing $k_2$. But the generation cost increases as the perturbation ratio decreases in both cases. When the branch susceptance is increased, the generation cost of perturbing $k_4$ almost remains invariant, while it increases with the perturbation ratio by perturbing $k_2$. The simulation result indicates that, by appropriately selecting the perturbed branch and the perturbation magnitude, we can reduce the increasing operation cost of MTD.

\begin{table}[htbp]
\centering
 \caption{Parameters of the generators}\label{table:opfgenerationcost}
 \begin{tabular}{c|ccccc}
  \hline
 Generation bus & 1  & 2 & 3 & 6 & 8\\
   \hline
 $P^g_{max}$ & 300  & 50 & 30 & 50 & 20\\
    \hline
 $\mu_i$(\$/MWh)& 20 & 30 & 40 & 50 & 35\\
    \hline
 \end{tabular}
\end{table}

\section{Conclusion}\label{section:conclusion}
In this paper, with the DC power flow model, we analyzed the completeness, deployment and the increasing operation cost of MTD in terms of thwarting stealthy FDI attacks constructed with old system information. To begin with, we proved that an MTD is complete to defeat all FDI attacks constructed with former branch parameters only if the number of branches $l$ is larger than or equal to twice that of the system states $n$ (i.e., $l \geq 2n$, where $n+1$ is the number of system buses), and the susceptances of more than $n$ branches, which cover all buses, are perturbed. Besides, we prove that we can never realize a complete MTD if the power transmission system has a bus that is only connected by a single branch. Further, we prove that the susceptance perturbation magnitude almost does not affect the dimension of the stealthy attack space after MTD. Based on this result, we presented guidance on effective MTD for minimizing the dimension of the stealthy stealthy attack space, maximizing the number of covered buese and reducing the operation cost. Finally, we illustrated and demonstrated our findings with the IEEE standard test power systems.

\appendices
\section{}\label{section:theoremfindineidne}
\indent\indent\emph{Proof of Proposition \ref{proposition:detectionoffdiandidentify}.} Since $\mathbf{H}^u$ can be linearly represented by $\mathbf{H}'$, we have $\mathbf{H}^u\textbf{c}_u \in S(\mathbf{H}')$ for any $\textbf{c}_u \in \mathbb{R}^u$. Suppose $\textbf{c}_v = \textbf{0}_{v}$. Then, we have $\textbf{a} = \mathbf{H}^u\textbf{c}_u + \mathbf{H}^v\textbf{c}_v \in S(\mathbf{H}')$ (i.e., $\textbf{a} \in \mathcal{A}_s$), which means that we cannot detect this FDI attack. Therefore, we must have $\textbf{c}_v \neq \textbf{0}_{v}$ for detecting the attack vector $\textbf{a}$.

\section{ }\label{section:propossitiongwg}
\indent\indent\emph{Proof of Proposition \ref{proposition:theoremvaluable}.} (Sufficiency) Since $\gamma = R([\mathbf{H}~\mathbf{H}']) = 2n$, we have $dim(\mathcal{A}_s)= dim(S(\mathbf{H})\cap S(\mathbf{H}'))= 0$. It follows that $\mathcal{A}_S  = \{\textbf{0}\}$. Therefore, the MTD is complete.

(Necessity) Since the MTD is complete, we have $\mathcal{A}_s = \{\textbf{0}\}$. It follows that $dim(\mathcal{A}_s) = dim(S(\mathbf{H})\cap S(\mathbf{H}'))= 2n -  R([\mathbf{H}~\mathbf{H}']) = 0$. Thus, we have $\gamma = R([\mathbf{H}~\mathbf{H}'])= 2n$.

\section{}\label{section:theorem43addad}
\indent\indent\emph{Proof of Theorem \ref{theorem:conditionforcompletedadsa}. } Considering the first condition, we resort to a contradiction. That is, suppose that the MTD is complete but $l < 2n$. Let $\widehat{\mathbf{H}}$ and $\widehat{\mathbf{H}}'$ denote the measurement matrices before and after MTD with a fully measured power system, respectively. Thus, with any partially measured power system, the measurement matrix $\mathbf{H}$ is formed by selecting some rows from $\widehat{\mathbf{H}}$. Therefore, the combined matrix $[\mathbf{H}~\mathbf{H}']$ can be formed by selecting some rows from the combined matrix $[\widehat{\mathbf{H}}~\widehat{\mathbf{H}}']$. It follows that $R([\mathbf{H}~\mathbf{H}']) \leq R([\widehat{\mathbf{H}}~\widehat{\mathbf{H}}'])$. We have proved that $R([\widehat{\mathbf{H}}~\widehat{\mathbf{H}}']) = 2n$ only if the number of branches $l$ is greater than or equal to $2n$. Therefore, we can derive that $R([\mathbf{H}~\mathbf{H}']) \leq R([\widehat{\mathbf{H}}~\widehat{\mathbf{H}}']) < 2n$ if $l < 2n$. This indicates that the MTD is not complete, which contradicts to the original assumption. Therefore, we must guarantee $l \geq 2n$ for achieving a complete MTD.

Next, we consider the second condition. Let $\mathcal{M}_{mtd}$ be the set of buses covered by the perturbed branches and $n_s$ is the size of $\mathcal{M}_{mtd}$. $\mathbf{H}^q$ is a submatrix formed by selecting $q$ columns from $\mathbf{H}'$ such that the combined matrix $[\mathbf{H}~\mathbf{H}^q]$ has full column rank. Let $\mathcal{M}^q_{mtd}$ be an index set of these $q$ columns in $\mathbf{H}'$. Since $\mathcal{M}^q_{mtd} \subseteq \mathcal{M}_{mtd}$, we have $q \leq n_s$. If the MTD is complete, then we have $q=n$. Therefore, $n_s \geq n$, that is, the perturbed branches must cover all buses.

\section{}\label{section:theorem44naifnsnfi}
\indent\indent\emph{Proof of Theorem \ref{theorem:neverscannot}.} Suppose $t$ is the bus that is only connected by a single branch. And the susceptance of the connected branch $k =\{t', t\}$ is $b_{t' t}$. Let $\textbf{e}_i \in \{0,1\}^n$ be a vector with a unit in the $i$th position and zero elsewhere, and let $\textbf{u}_{ij} = \textbf{e}_i - \textbf{e}_j$ ($k=\{i,j\} \in \mathcal{L}$) be a vector with ``1" in the $i$th position and ``-1" in the $j$th position and zero elsewhere. Considering the fully measured case, let the $t$th column of the branch-bus shift factor matrix $\mathbf{S}$ and the symmetric admittance matrix $\mathbf{B}$ be $\textbf{s}_t$ and $\textbf{b}_t$, respectively. Then, we have $\textbf{s}_t = -b_{t' t} \textbf{e}_t$ and $\textbf{b}_t = b_{t' t} \textbf{u}_{t't}$. Suppose $\textbf{s}'_t = -b'_{t' t} \textbf{e}_t$ and $\textbf{b}'_t = b'_{t' t} \textbf{u}_{t't}$ after MTD. Then, we can derive that $\textbf{s}_t = \frac{b_{t' t}}{ b'_{t' t}}\textbf{s}'_t$ and $\textbf{b}_t = \frac{b_{t' t}}{ b'_{t' t}}\textbf{b}'_t$. Let $\textbf{h}_t$ and $\textbf{h}'_t$ be the $t$th column of $\mathbf{H}$ and $\mathbf{H}'$ before and after MTD, respectively. Then, we have $\textbf{h}_t = \frac{b_{t' t}}{ b'_{t' t}} \textbf{h}'_t$. Since any $\mathbf{H}$ can be formed by selecting some rows from the measurement matrix under the fully measured case, we can derive $\textbf{h}_t = \frac{b_{t' t}}{ b'_{t' t}} \textbf{h}'_t$ under the partially measured case as well. It follows that $S(\mathbf{H}') = S(\mathbf{H})$ if we only perturb the branch $k$. Therefore, we can never obtain $\gamma = R([\mathbf{H}~\mathbf{H}']) = 2n$.

\section{}\label{section:propositiondadau}
\indent\indent\emph{Proof of Proposition \ref{proposition:sparsedeltaHdada}.} Let $\textbf{e}_i \in \{0,1\}^n$ be a vector with a unit in the $i$th position and zeros elsewhere, and let $\textbf{u}_{ij} = \textbf{e}_i - \textbf{e}_j$ ($k=\{i,j\} \in \mathcal{L}$) be a vector with ``1" in the $i$th position and ``-1" in the $j$th position and zeros elsewhere. Then, we can rewrite $\mathbf{A}$ and $\mathbf{D}$ as

\begin{equation}\label{rewritematrices}
  \mathbf{A} = \sum^{}_{\overset{k \in \mathcal{L}}{ k=\{i,j\}}} \textbf{e}_k \textbf{u}^T_{ij}~~,~~~~~~~~~ \mathbf{D} = \sum^{}_{\overset{k \in \mathcal{L}}{ k=\{i,j\}}} -b_{ij}\textbf{e}_k \textbf{e}^T_k.
\end{equation}

Supposing the diagonal branch susceptance matrix after MTD is $\mathbf{D}'$ and $\Delta \mathbf{D} = \mathbf{D}' - \mathbf{D}$, then we can derive $\Delta \mathbf{D}$ as
\begin{equation}\label{thedifferencesusceptance}
\begin{aligned}
  \Delta \mathbf{D}  & = \sum^{}_{\overset{k_d \in \mathcal{L}_D}{ k_d=\{i_d,j_d\}}} -(b'_{i_dj_d}- b_{i_dj_d})\textbf{e}_{k_d}\textbf{e}^T_{k_d}\\&=\sum^{}_{\overset{k_d \in \mathcal{L}_D}{ k_d=\{i_d,j_d\}}} -\Delta b_{i_d j_d}\textbf{e}_{k_d}\textbf{e}^T_{k_d}.
\end{aligned}
\end{equation}
We can see that there are non-zero elements in positions $(k_d, k_d)$ with $k_d \in \mathcal{L}_D$ and zeros elsewhere in $\Delta \mathbf{D}$. Since $\textbf{e}^T_i\textbf{e}_j = 0$ if $i \neq j$ and $\textbf{e}^T_i\textbf{e}_j = 1$ if $i = j$, we can derive that the change of the branch-bus shift factor matrix is
\begin{equation}\label{branchbusshiftfactormatrix}
\begin{aligned}
  \Delta \mathbf{S} = \Delta \mathbf{D} \mathbf{A} & = \sum^{}_{\overset{k_d \in \mathcal{L}_D}{k_d=\{i_d,j_d\}}} -(\Delta b_{i_d j_d} \textbf{e}_{k_d}\textbf{e}^T_{k_d})\big(\textbf{e}_{k_d} \textbf{u}^T_{i_dj_d}\big)\\
  &=\sum^{}_{\overset{k_d \in \mathcal{L}_D}{k_d=\{i_d,j_d\}}} -\Delta b_{i_d j_d} \textbf{e}_{k_d}\textbf{u}^T_{i_dj_d}
\end{aligned}
\end{equation}
We can observe that $\Delta \mathbf{S}$ is a sparse matrix with the $k_d$th row is non-zero and other rows are all zero. That is, if $i \in \mathcal{L}_D$, then the $i$th row of $\Delta \mathbf{S}$ is $\Delta \mathbf{S}_i = [\cdots -\Delta b_{i_dj_d} \cdots \Delta b_{i_dj_d} \cdots]$ with $-\Delta b_{i_dj_d}$ in the $i$th column and $\Delta b_{i_dj_d}$ in the $j$th column; if $i \notin \mathcal{L}_D$, $\Delta \mathbf{S}_i =[0\cdots 0 \cdots 0]$. Similarly, since $\mathbf{B} = \mathbf{A}^T\mathbf{S}$, we have $\Delta \mathbf{B} = \mathbf{A}^T \Delta \mathbf{S}$
\begin{equation}\label{branchbusshiftfactormatrix}
\begin{aligned}
  \Delta \mathbf{B} = \mathbf{A}^T \Delta \mathbf{S} & = \sum^{}_{\overset{k_d \in \mathcal{L}_D}{k_d=\{i_d,j_d\}}} -\Delta b_{i_d j_d} (\textbf{u}_{i_d j_d} \textbf{e}^T_{k_d})\big(\textbf{e}_{k_d} \textbf{u}^T_{i_dj_d}\big)\\
  &=\sum^{}_{\overset{k_d \in \mathcal{L}_D}{k_d=\{i_d,j_d\}}} -\Delta b_{i_d j_d} \textbf{u}_{i_d j_d} \textbf{u}^T_{i_dj_d}.
\end{aligned}
\end{equation}
$\Delta \mathbf{B}$ is a sparse matrix with the $i_d$th and $j_d$th columns are non-zero and the other columns are all zero.

On the basis of the measurement matrix $\mathbf{H}$ (before MTD) and $\mathbf{H}'$ (after MTD). Since $\mathbf{H}' = [\mathbf{B} + \Delta \mathbf{B}; \mathbf{S} + \Delta \mathbf{S}; -\mathbf{S}- \Delta \mathbf{S}]$,
we can derive that the difference of the measurement matrix is
\begin{equation}\label{measurementmatrix2}
  \Delta \mathbf{H} = \mathbf{H}' -  \mathbf{H} =
  \begin{bmatrix}
  \Delta \mathbf{B} \\
  \Delta \mathbf{S}  \\
  - \Delta \mathbf{S}
  \end{bmatrix}.
\end{equation}
Obviously, we can see that $\Delta \mathbf{H}$ is a sparse matrix with non-zero elements in the $i_d$th and $j_d$th column with $k_d=\{i_d, j_d\} \in \mathcal{L}_D$.

For example, if we only perturb a single branch, that is, $\mathcal{L}_d=\{k_d\}$ with $k_d = \{i_d, j_d\}$, then $\Delta \mathbf{H}$ under a fully measured case is stated as
\begin{small}
\begin{equation}\label{sparsedeltahder}
  \Delta \mathbf{H}=
  \!\!\begin{bmatrix}
  \mathbf{0} & \vdots &  \mathbf{0} & \vdots & \mathbf{0} & {} \\
  \cdots & -\Delta b_{i_dj_d}  &  \cdots & \Delta b_{i_dj_d} & \cdots & {\}i_d} \\
  \mathbf{0} & \vdots  &  \mathbf{0} & \vdots & \mathbf{0} & {} \\
  \cdots &  \Delta b_{i_dj_d} &  \cdots & -\Delta b_{i_dj_d}& \cdots & {\} j_d} \\
  \mathbf{0} &  \vdots & \mathbf{0} & \vdots & \mathbf{0} & {} \\
  \cdots & -\Delta b_{i_dj_d}  &  \cdots & \Delta b_{i_dj_d} & \cdots & {\}n + k_d} \\
  \mathbf{0} & \vdots  &  \mathbf{0} & \vdots & \mathbf{0} & {} \\
  \cdots &  \Delta b_{i_dj_d} &  \cdots & -\Delta b_{i_dj_d}& \cdots & {\}n + l+ k_d} \\
  \mathbf{0} &  \vdots & \mathbf{0} & \vdots & \mathbf{0} & {} \\
  {} &  \overset{\underbrace{}}{i_d} & {} & \overset{\underbrace{}}{j_d} & {} & {} \\
  \end{bmatrix}\!\!,
\end{equation}
\end{small}where $\Delta b_{i_dj_d}$ is the susceptance perturbation of branch $k_d$, $\mathbf{0}$ is a zero matrix/vector. We can see that all non-zero elements in $\Delta \mathbf{H}$ are in the $i_d$th column and the $j_d$th column. Since any measurement matrix is formed by selecting some rows from the measurement matrix $\mathbf{H}$ under the fully measured case, we can derive that $\Delta \mathbf{H}$ is also a sparse matrix with non-zero elements in the $i_d$th and $j_d$th column with $k_d=\{i_d, j_d\} \in \mathcal{L}_D$.

\section{}\label{section:propostionsdsfdad}
\indent\indent\emph{Proof of Proposition \ref{propostion:pppp}.} We have $\mathbf{C}' = [\mathbf{H}~\mathbf{H}'] + [\mathbf{0}_{m\times n} ~ \Delta \mathbf{H}] = \mathbf{C} + [\mathbf{0}_{m\times n} ~ \Delta \mathbf{H}] $, where $\mathbf{0}_{m\times n}$ is an $m$ by $n$ zero matrix, $\Delta \mathbf{H}$ is a matrix corresponding to the susceptance perturbation of the new branch. According to the equation (\ref{sparsedeltahder}), $R(\Delta \mathbf{H}) = 1$ when one more branch is perturbed. Since $R(\mathbf{C}') \leq R(\mathbf{C}) + R( \Delta \mathbf{H})$, we have $R(\mathbf{C}') \leq R(\mathbf{C}) + 1$. Conversely, we have $\mathbf{C} = [\mathbf{H}~\mathbf{H}'] + [\mathbf{0}_{m\times n} ~ -\Delta \mathbf{H}] = \mathbf{C}' + [\mathbf{0}_{m\times n} ~ -\Delta \mathbf{H}]$. Thus, we can derive that $R(\mathbf{C}) \leq R(\mathbf{C}') + R( -\Delta \mathbf{H}) \Rightarrow R(\mathbf{C}') \geq R(\mathbf{C})  - 1$. Therefore, we have $R(\mathbf{C})  - 1 \leq R(\mathbf{C}')\leq R(\mathbf{C}) + 1 $.

\section{ }\label{section:propsoitiond47nins}
\indent\indent\emph{Proof of Proposition \ref{proposition:proforrankincreasingcase1}.} Here we only analyze the case when $i_d \notin \mathcal{M}^q_{mtd}$, because we can draw a same conclusion when $j_d \notin \mathcal{M}^q_{mtd}$. We denote det$(\cdot)$ as the determinant of a matrix. $\textbf{a}_i$ is the $i$th column of a matrix $\mathbf{A}$. Since $i_d \notin \mathcal{M}^q_{mtd}$ and $j_d \notin \mathcal{M}^q_{mtd}$, the $i_d$th column $\textbf{h}'_{i_d}$ and the $j_d$th column $\textbf{h}'_{j_d}$ of $\mathbf{H}'$ can be linearly represented by the columns in $[\mathbf{H}~\mathbf{H}^q]$. It follows that $R([\mathbf{H}~\mathbf{H}^q~\textbf{h}'_{i_d}]) = R([\mathbf{H}~\mathbf{H}^q]) = R([\mathbf{H}~\mathbf{H}'])$. After the susceptance perturbation of branch $k_d$, $\textbf{h}'_{i_d}$ and $\textbf{h}'_{j_d}$ respectively become $\textbf{h}''_{i_d}$ and $\textbf{h}''_{j_d}$ according to the proof of Proposition \ref{proposition:sparsedeltaHdada}. We rearrange the matrix $\mathbf{C}'$ as $\mathbf{C}' = [\mathbf{H}~\mathbf{H}^q~\textbf{h}''_{i_d}~\textbf{h}''_{j_d}~\mathbf{H}''_{r}]$, where $\mathbf{H}''_{r}$ is a submatrix containing the columns in $\mathbf{H}^{p}$ ($\mathbf{H}' = [\mathbf{H}^q ~ \mathbf{H}^{p}]$) that are not changed after the perturbation of branch $k_d$. Let $\mathbf{C}'' = [\mathbf{H}~\mathbf{H}^q~\textbf{h}'_{i_d}~\textbf{h}''_{i_d}~\textbf{h}''_{j_d}~\mathbf{H}''_{r}]$ be a matrix by filling the column $\textbf{h}'_{i_d}$ into $\mathbf{C}'$. Then, we can derive that $R(\mathbf{C}'') = R(\mathbf{C}')$.

We calculate $R(\mathbf{C}'')$ through elementary column operations. Since only the element relating to branch $k_d$ is changed in column $\textbf{h}''_{i_d}$, by subtracting $\textbf{h}''_{i_d}$ from $\textbf{h}'_{i_d}$, we have $\Delta \textbf{h}''_{i_d} =  \textbf{h}''_{i_d} - \textbf{h}'_{i_d} = [0~\cdots~0~ \pm (\lambda - 1)b_{i_d j_d}~0~\cdots~ 0~ \pm (\lambda - 1)b_{i_d j_d}~0~\cdots~0]^T$, where $ (\lambda - 1)b_{i_d j_d}$ and $-(\lambda - 1)b_{i_d j_d}$ are the only non-zero elements in the vector $\Delta \textbf{h}''_{i_d}$. The number of non-zero elements in this column depends on the measured case of the power system. But it is less than 4 according to the example given in the equation (\ref{sparsedeltahder}). For the elementary column operation, we transfer the column $\textbf{h}'_{i_d}$ to $\Delta \textbf{h}''_{i_d}$ in $\mathbf{C}''$. Since the other elements of the column $\Delta \textbf{h}''_{i_d}$ are zero, any other column operations on this column are only related to $(\lambda - 1)b_{i_d j_d}$. Therefore, it is equivalent to operate on $(\lambda - 1)b_{i_d j_d}$ for elementary column operations for any $\lambda$ if $\lambda \neq 1$. For example, we perturb the susceptance of branch $k_d$ to $\lambda_1 b_{i_dj_d}$ and $\lambda_2 b_{i_dj_d}$, respectively. Assume that only the branches are monitored by meters and each branch is monitored by one meter. Then, we can derive that
\begin{equation}\label{hpie}
\begin{aligned}
  &\Delta\textbf{h}''_{i_d} = \![0~\cdots~0~( 1 -\lambda)b_{i_d j_d}~0~\cdots~ 0]^T\!, \\
   &~~~~~~~~~~~~~~~~~~\overset{\underbrace{}}{k_dth~element}~~~~~~~~~~~~
\end{aligned}
\end{equation}
If the required column operation is to multiply $\Delta \textbf{h}''_{i_d}$ with a constant $\eta$, then the result is $\textbf{h} = \eta \Delta \textbf{h}''_{i_d}$. We find $\textbf{h} = \eta \![0~\cdots~0~( 1 -\lambda_1)b_{i_d j_d}~0~\cdots~ 0]^T\! =\eta \times \frac{\lambda_1}{\lambda_2} \![0~\cdots~0~( 1 -\lambda_2)b_{i_d j_d}~0~\cdots~ 0]^T\!$. That is, we can always obtain the same column operation result $\textbf{h}$ regardless of the value of $\lambda$.

Therefore, the impact of $\Delta\textbf{h}''_{i_d}$ on $R(\mathbf{C}'')$ is not related to the value of $\lambda$. Thus, the impact of the column $\textbf{h}''_{i_d}$ on $R(\mathbf{C}'')$ is not related to the value of $\lambda$. Since $R(\mathbf{C}'') = R(\mathbf{C}')$, the impact of the column $\textbf{h}''_{i_d}$ on $R(\mathbf{C}')$ is not related to the value of $\lambda$. As for the column $\textbf{h}''_{j_d}$, we can draw the same conclusion by using elementary column operations. Therefore, $\Delta \gamma = R(\mathbf{C}') - R(\mathbf{C})$ does not change with $\lambda$ ($\lambda > 0$ and $\lambda \neq 1$).

\section{}\label{section:proposition48dsdfa}
\indent\indent\emph{Proof of Proposition \ref{proposition:decrease} .} We prove this proposition in \emph{\textbf{Case 2}} and \emph{\textbf{Case 3}}, respectively. We denote det$(\cdot)$ as the determinant of a matrix. $\textbf{a}_i$ is the $i$th column of a matrix $\mathbf{A}$.

\underline{\emph{\textbf{Case 2: $i_d \in \mathcal{M}^q_{mtd}$ and $j_d \notin \mathcal{M}^q_{mtd}$}}}

Since $\Delta\gamma = -1$, we have $R([\mathbf{H}~\mathbf{H}'']) = R([\mathbf{H}~\mathbf{H}']) - 1$. It follows that we must have $R([\mathbf{H}~\mathbf{H}^q_{r} ~ \textbf{h}''_{i_d}]) = R([\mathbf{H}~\mathbf{H}^q]) - 1$ when $\lambda = \lambda^*$, where $\mathbf{H}^q_{r}$ is a submatrix formed by deleting the column $\textbf{h}'_{i_d}$ from $\mathbf{H}^q$, $\textbf{h}'_{i_d}$ is the $i_d$th column of $\mathbf{H}'$, $\textbf{h}''_{i_d}$ is the $i_d$th column of $\mathbf{H}''$ after the perturbation of branch $k_d$. If we select any $n+q$ rows that contain the perturbed rows (i.e., there are elements $\lambda b_{i_d j_d}$ and/or $-\lambda b_{i_d j_d}$ in this row) and any $n+q$ columns from the combined matrix $[\mathbf{H}~\mathbf{H}^q_{r} ~ \textbf{h}''_{i_d}]$, we can form an $n+q$ by $n+q$ square matrix $\mathbf{Q}$. Using the Leibniz formula, det$(\mathbf{Q})$ is a linear function of $\lambda b_{i_d j_d}$, that is, det$(\mathbf{Q}) = a  b_{i_d j_d} \lambda + b$, where $a$ and $b$ are values calculated by other elements. Specifically, $a$ and $b$ are different if we choose different rows and columns to form $\mathbf{Q}$. Because the combined matrix $[\mathbf{H}~\mathbf{H}^q_{r} ~ \textbf{h}''_{i_d}]$ has full column rank when $\lambda = 1$, there exists $\mathbf{Q}$ such that det$(\mathbf{Q}) = a  b_{i_d j_d} + b \neq 0$. Therefore, $a$ and $b$ cannot be 0 at the same time. If $a = 0$, then det$(\mathbf{Q}) = a b_{i_d j_d} \lambda + b \neq 0$ for any $\lambda$. Therefore, there does not exist $\lambda$ such that det($\mathbf{Q}$)$= 0$. If $a \neq 0$, then det$(\mathbf{Q}) = a  b_{i_d j_d} \lambda + b = 0$ has an unique solution $\frac{-b}{a  b_{i_d j_d}}$. As $\Delta\gamma = -1$ when $\lambda = \lambda^*$, we have $R([\mathbf{H}~\mathbf{H}^q_{r} ~ \textbf{h}''_{i_d}]) = n+q -1$. It follows that, for any $\mathbf{Q}$, we have det($\mathbf{Q}$)$=  a  b_{i_d j_d} \lambda^* + b =0$. This indicates that $\lambda^*$ must be equal to $\frac{-b}{a  b_{i_d j_d}}$. Therefore, $\Delta \gamma = -1$ only if there exists $\lambda^*$ such that $\lambda_{min} \leq \lambda^* \leq \lambda_{max}$ and $\lambda^*  \neq 1$ and $\lambda = \lambda^*$.

\underline{\emph{\textbf{Case 3: $i_d \in \mathcal{M}^q_{mtd}$ and $j_d \in \mathcal{M}^q_{mtd}$}}}

After branch $k_d$ is perturbed, we can rearrange $\mathbf{C}' = [\mathbf{H}~\mathbf{H}'']$ as $\mathbf{C}' = [\mathbf{H}~\mathbf{H}^q_{r}~\textbf{h}''_{i_d}~\textbf{h}''_{j_d}~ \mathbf{H}^{p}]$, where $\mathbf{H}^q_{r}$ is a matrix containing the rest columns of $\mathbf{H}^q$ that are not changed, $\textbf{h}''_{i_d}$ and $\textbf{h}''_{j_d}$ are the $i_d$th column and $j_d$th column in $\mathbf{H}''$ that are perturbed. If we select any $n+q$ rows that contain the perturbed rows (i.e., there are elements $\lambda b_{i_d j_d}$ and/or $-\lambda b_{i_d j_d}$ in this row) and any $n+q$ columns from the combined matrix $[\mathbf{H}~\mathbf{H}^q_{r}~\textbf{h}''_{i_d}~\textbf{h}''_{j_d}]$, we can from an $n+q$ by $n+q$ square matrix $\mathbf{Q}$. Using the Leibniz formula, the determinant of $\mathbf{Q}$ is det$(\mathbf{Q}) = a  b_{i_d j_d} \lambda + b$, where $a$ and $b$ are coefficients after the calculation. As there exists $\mathbf{Q}$ such that we have det$(\mathbf{Q}) = a  b_{i_d j_d} + b \neq 0$ when $\lambda = 1$, $a$ and $b$ cannot be 0 at the same time. Then, if $a = 0$, there exists $\mathbf{Q}$ such that det$(\mathbf{Q})\neq 0$ for any $\lambda$. Therefore, $\Delta \gamma \neq -1$ for any $\lambda$. If $a\neq 0$, det$(\mathbf{Q}) =a  b_{i_d j_d} \lambda+ b = 0$ has an unique solution $\frac{-b}{a  b_{i_d j_d}}$. Thus, if $R(\mathbf{Q}) = n + q -1$ (i.e., $\Delta \gamma = -1$) when $\lambda = \lambda^*$, then $\lambda^*$ must be equal to $\frac{-b}{a  b_{i_d j_d}}$. Therefore, $\Delta \gamma = -1$ only if there exists $\lambda^*$ such that $\lambda_{min} \leq \lambda^* \leq \lambda_{max}$ and $\lambda^* \neq 1$ and $\lambda = \lambda^*$.

\section{}\label{section:theorem49bbuda}
\indent\indent\emph{Proof of Theorem \ref{theorem:increasetheorem}.}

\underline{\emph{\textbf{Case 1: $i_d \notin \mathcal{M}^q_{mtd}$ and $j_d \notin \mathcal{M}^q_{mtd}$}}}

In \emph{\textbf{Case 1}}, we have proved that the susceptance-perturbation magnitude does not affect the change of security factor $\gamma$. Therefore, if there exists $\lambda = \lambda^*$ such that $\Delta \gamma = 1$, then this result will not change if we change the perturbation magnitude to the other values. The proof of necessity is obvious.

\underline{\emph{\textbf{Case 2: $i_d \in \mathcal{M}^q_{mtd}$ and $j_d \notin \mathcal{M}^q_{mtd}$}}}

(Sufficiency) Suppose there exists a column $\textbf{h}^{p}_{t}$ in $\mathbf{H}^{p}$ such that $\textbf{h}^{p}_{t}$ cannot be linearly represented by the columns in $[\mathbf{H}~\mathbf{H}^q_{r}]$ when we perturb the susceptance of branch $k_d$ to $\lambda^* b_{i_d j_d}$, where $\mathbf{H}^q_{r}$ is a submatrix formed by the columns of $\mathbf{H}^q$ that are not changed after the perturbation. Since $R([\mathbf{H} ~ \mathbf{H}^q]) = R([\mathbf{H}~\mathbf{H}'])$, we cannot find another two columns in $\mathbf{H}^{p}$ that cannot be linearly represented by the columns in $[\mathbf{H}~\mathbf{H}^q_r]$. Therefore, in order to have $\Delta \gamma = R([\mathbf{H}~\mathbf{H}''])- R([\mathbf{H}~\mathbf{H}']) = 1$, at least one of $\textbf{h}''_{i_d}$ and $\textbf{h}''_{j_d}$ cannot be linearly represented by the columns in $[\mathbf{H}~\mathbf{H}^q_{r}]$, where $\textbf{h}''_{i_d}$ and $\textbf{h}''_{j_d}$ are respective the $i_d$th column and $j_d$th column in $\mathbf{H}''$ after the perturbation of branch $k_d$. Here, we suppose that $\textbf{h}''_{i_d}$ cannot be linearly represented by the columns in $[\mathbf{H}~\mathbf{H}^q_r]$ (we can draw the same conclusion if $\textbf{h}''_{j_d}$ cannot be linearly represented by the columns in $[\mathbf{H}~\mathbf{H}^q_r]$). Let $\mathbf{Q} = [\mathbf{H}~\mathbf{H}^q_{r}~\textbf{h}^{p}_{t}~\textbf{h}''_{i_d}]$ be a matrix of full column rank and $R(\mathbf{Q}) = R(\mathbf{H}~\mathbf{H}'']) = R([\mathbf{H}~\mathbf{H}']) + 1$. We select $ (n+q+1)$ rows that contain the perturbed rows (i.e., there are elements $\lambda^* b_{i_d j_d}$ and/or $-\lambda^* b_{i_d j_d}$ in this row) and $ (n+q+1)$ columns from $\mathbf{Q}$ to form an $ (n+q+1)$ by $(n+q+1)$ square matrix $\mathbf{\bar{Q}}$. Using the Leibniz formula, we can derive that det$(\mathbf{\bar{Q}})$ is a linear function of $\lambda^* b_{i_d j_d}$, that is, det$(\mathbf{\bar{Q}}) = a  b_{i_d j_d} \lambda^* + b$, where $a$ and $b$ are values after the calculation. Specifically, $a$ and $b$ are different if we choose different rows and columns in $\mathbf{Q}$ to form $\mathbf{\bar{Q}}$. Since $\Delta \gamma = 1$, we can derive that there exists $\mathbf{\bar{Q}}$ such that det$(\mathbf{\bar{Q}}) = a  b_{i_d j_d} \lambda^*  + b \neq 0$. Therefore, $a$ and $b$ cannot be 0 at the same time.

If $a =0$, we have $b\neq 0$. Therefore, for any $\lambda^*$, we have det$(\mathbf{\bar{Q}}) = a  b_{i_d j_d} \lambda + b \neq 0$. It follows that we always have $\Delta \gamma = 1$ for $\lambda > 0$ ($\lambda \neq 1$). If $a\neq0$, we exploit whether or not there exists $\lambda$ such that $\Delta \gamma < 1 $. If $\Delta \gamma < 1 $, we must have $R(\mathbf{Q})\leq R([\mathbf{H}~\mathbf{H}'])$. It indicates that, for any $\mathbf{\bar{Q}}$, we have det$(\mathbf{\bar{Q}}) = 0$. As we know, $\lambda^* = 1$ is the unique solution for det$(\mathbf{\bar{Q}}) = a  b_{i_d j_d} \lambda^* + b = 0$ (for any $\mathbf{\bar{Q}}$). However, since $\lambda^* \neq 1$, we can never have det$(\mathbf{\bar{Q}}) =0 $ for any $\mathbf{\bar{Q}}$. Therefore, $\Delta \gamma = 1$ for any $\lambda > 0$ ($\lambda \neq 1$).

(Necessity) Obviously, if $\Delta \gamma = 1$ for any $\lambda >0 $ ($\lambda \neq 1$), that there must exists $\lambda = \lambda^*$ such that $\Delta \gamma = 1$.

\underline{\emph{\textbf{Case 3: $i_d \in \mathcal{M}^q_{mtd}$ and $j_d \in \mathcal{M}^q_{mtd}$}}}

(Sufficiency) Let $\mathbf{P} =  [\mathbf{H}~\mathbf{H}^q_{r}]$, where $\mathbf{H}^q_{r}$ is a submatrix containing the columns in $\mathbf{H}^q$ that are not changed after perturbing the susceptance of branch $k_d$ to $\lambda^* b_{i_d j_d}$. Because $R([\mathbf{H}~\mathbf{H}^q]) = R([\mathbf{H}~\mathbf{H}']) = n+q$, we can never find another three columns in $\mathbf{H}^{p}$ to form a combined matrix with $\mathbf{P}$ that has rank $n+q+1$. Therefore, supposing $R([\mathbf{H}~\mathbf{H}^q_{r}~\textbf{h}''_t~\textbf{h}''_{t'}~\textbf{h}''_{t''}])=n+q+1$, then at least one column in $\textbf{h}''_t,\textbf{h}''_{t'},\textbf{h}''_{t''}$ is either $\textbf{h}''_{i_d}$ or $\textbf{h}''_{j_d}$, where $\textbf{h}''_{i_d}$ and $\textbf{h}''_{j_d}$ are the columns being perturbed in $\mathbf{H}^q$ after the perturbation of branch $k_d$. Here we assume that is $\textbf{h}''_{i_d}$ (we can draw the same conclusion if it is $\textbf{h}''_{j_d}$). Let $\mathbf{Q}= [\mathbf{H}~\mathbf{H}^q_{r}~\textbf{h}''_{i_d}~\textbf{h}''_{t}~\textbf{h}''_{t'}]$. It follows that $R(\mathbf{Q}) = n+q+1$. We select $(n+q+1)$ rows that contain the perturbed rows (i.e., there are elements $\lambda^* b_{i_d j_d}$ and/or $-\lambda^* b_{i_d j_d}$ in this row) and $ (n+q+1)$ columns from $\mathbf{Q}$ to form an $ (n+q+1)$ by $(n+q+1)$ square matrix $\mathbf{\bar{Q}}$. Using the Leibniz formula, we can obtain that the determinant of $\mathbf{\bar{Q}}$ is det$(\mathbf{\bar{Q}}) = a  b_{i_d j_d} \lambda^* + b$, where $a$ and $b$ are coefficients after the calculation. Since there exists $\mathbf{\bar{Q}}$ such that det$(\mathbf{\bar{Q}}) = a  b_{i_d j_d}\lambda^* + b \neq 0$, $a$ and $b$ cannot be 0 at the same time.

If $a = 0$, we have $b \neq 0$. Thus, det$(\mathbf{\bar{Q}})\neq 0$ for any $\lambda^* > 0$. Therefore, $\Delta \gamma = 1$ for any $\lambda^* > 0$. If $a \neq 0$, as we know, $\lambda = 1$ is the unique solution for det$(\mathbf{\bar{Q}}) =a  b_{i_d j_d} \lambda^* + b = 0$. However, since $\lambda^* \neq 1$, we can never have det$(\mathbf{\bar{Q}}) =a  b_{i_d j_d} \lambda^*+ b \neq 0$ for any $\mathbf{\bar{Q}}$ and any $\lambda > 0$ ($\lambda^* \neq 1$). Therefore, $\Delta \gamma = 1$ for any $\lambda > 0$ ($\lambda \neq 1$).

(Necessity) Since $ \Delta \gamma = 1$ for any $\lambda >0 $ ($\lambda \neq 1$), we must have $\lambda = \lambda^*$ such that $\Delta \gamma = 1$.
\begin{scriptsize}

\end{scriptsize}
\vskip -2\baselineskip plus -1fil
\begin{IEEEbiography}[{\includegraphics[width=1in,height=1.25in,clip,keepaspectratio]{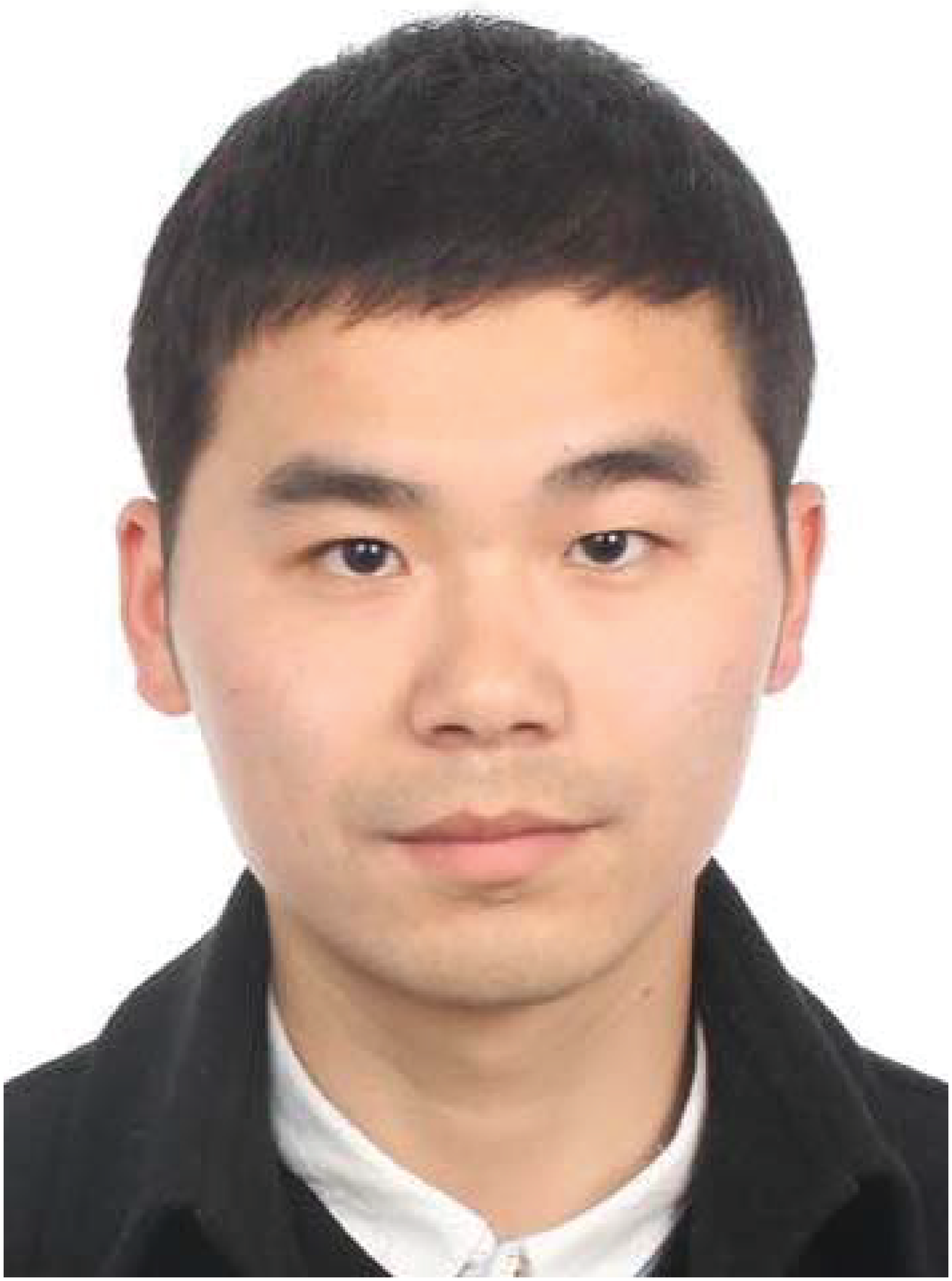}}] {Zhenyong Zhang}
is a fourth year Ph.D. candidate in the school of control science and engineering, Zhejiang University, Hangzhou, China. He received his bachelor degree in control science and engineering from Central South University, Changsha, China, in 2015. He is currently a visiting Ph.D. student at Singapore University of Technology and Design. His research interests include mobile computing and cyber-physical system security.
\end{IEEEbiography}
\vskip -2\baselineskip plus -1fil
\begin{IEEEbiography}[{\includegraphics[width=1in,height=1.25in,clip,keepaspectratio]{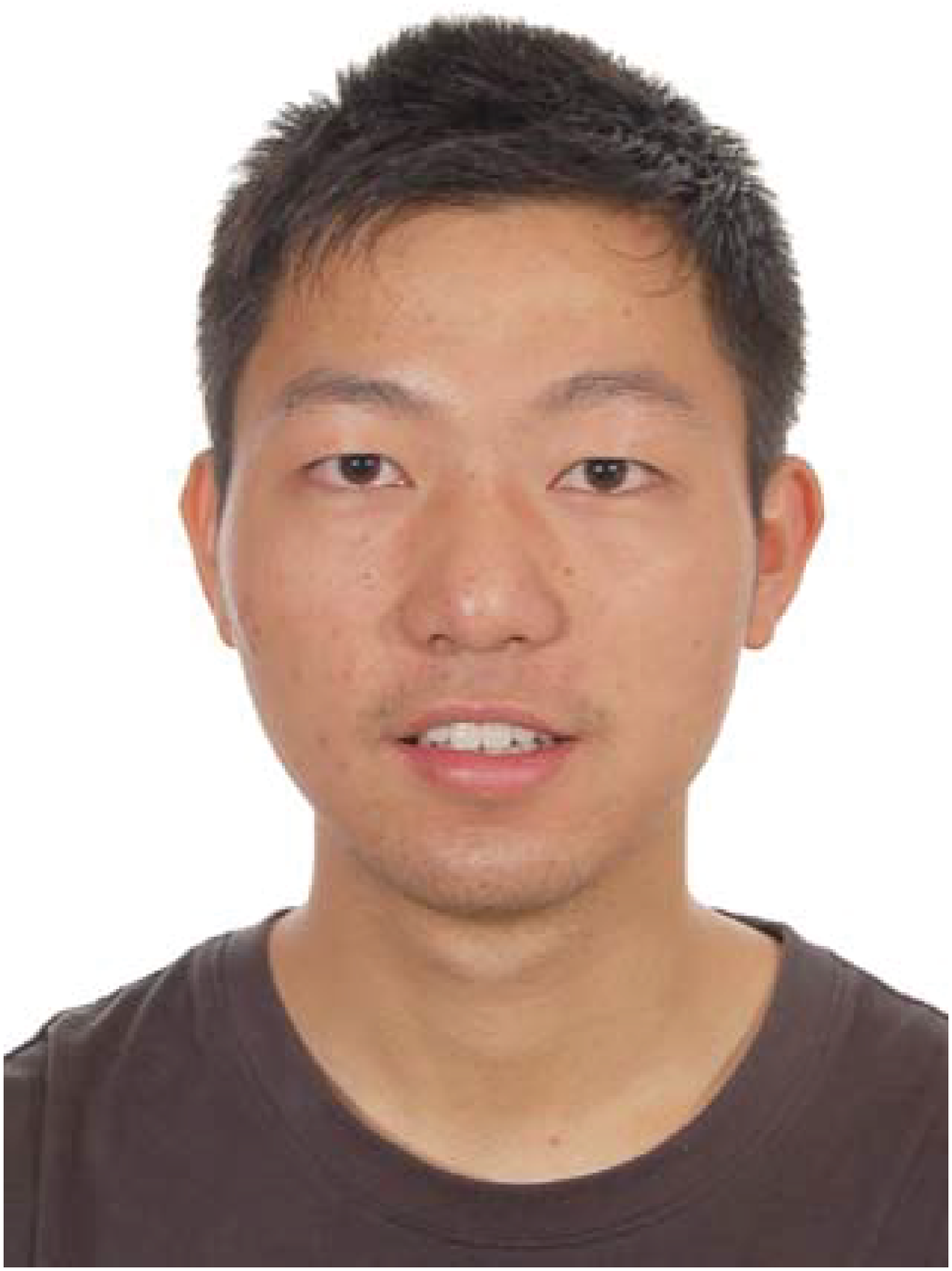}}] {Ruilong Deng} (S'11-M'14) received the B.Sc. and Ph.D. degrees both in Control Science and Engineering from Zhejiang University, Hangzhou, Zhejiang, China, in 2009 and 2014, respectively. He was a Research Fellow with Nanyang Technological University, Singapore, from 2014 to 2015; and an AITF Postdoctoral Fellow with the University of Alberta, Edmonton, AB, Canada, from 2015 to 2018. Currently, he is an Assistant Professor with the School of Computer Science and Engineering, Nanyang Technological University, Singapore. His research interests include smart grid, cyber security, and wireless networking.
\vspace{-0.95cm}
\end{IEEEbiography}
\begin{IEEEbiography}[{\includegraphics[width=1in,height=1.25in,clip,keepaspectratio]{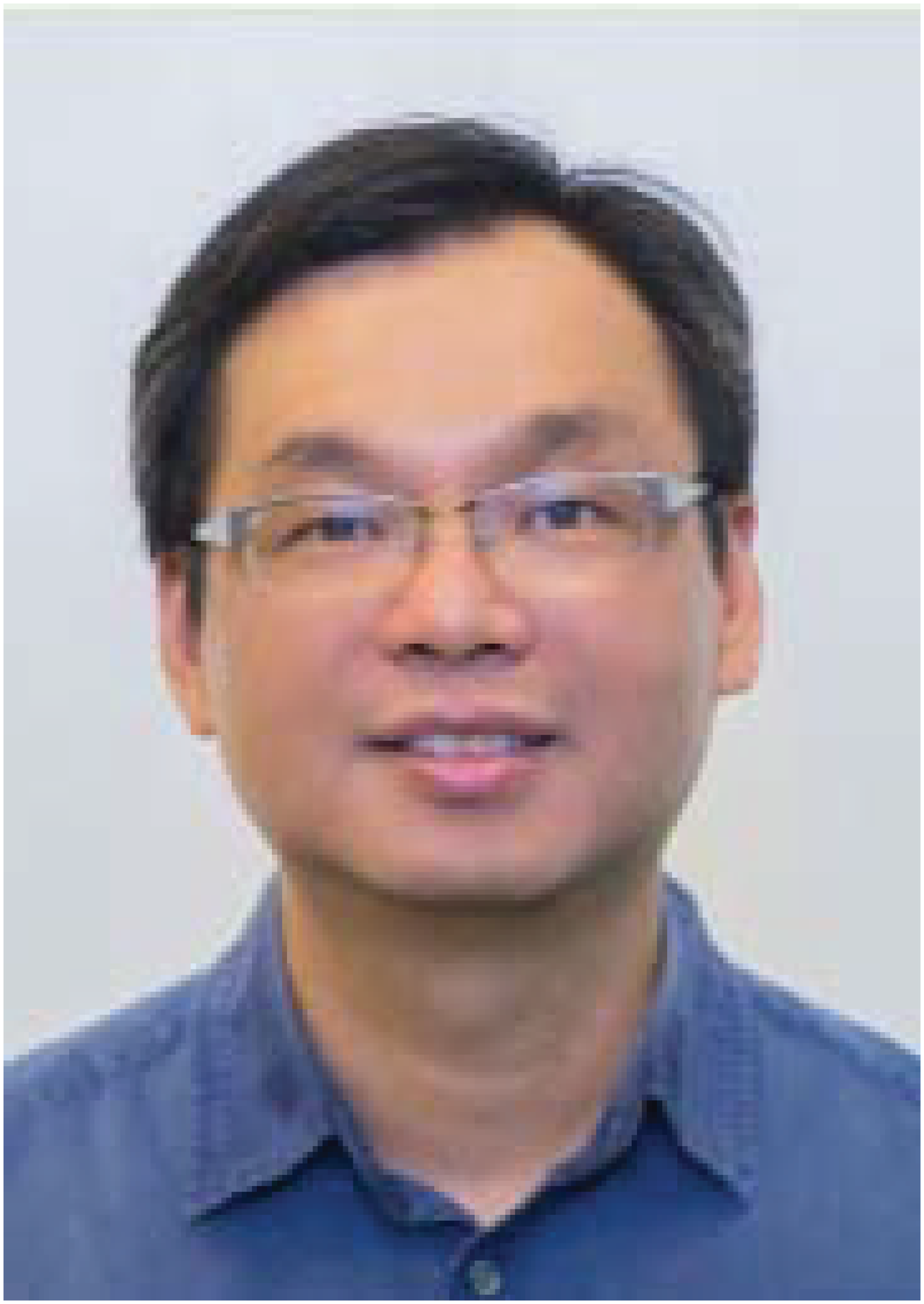}}] {David K. Y. Yau} (M'10-SM'13) received the B.Sc. from the Chinese University of Hong Kong, and M.S. and Ph.D. from the University of Texas at Austin, all in computer science. He has been Professor at Singapore University of Technology and Design since 2013. Since 2010, he has been Distinguished Scientist at the Advanced Digital Science Centre, Singapore. He was Associate Professor of Computer Science at Purdue University (West Lafayette). His research interests include cyber-physical system and network security/privcay, wireless sensor networks, and smart grid IT. He is a senior member of IEEE.
\vspace{-0.85cm}
\end{IEEEbiography}
\begin{IEEEbiography}[{\includegraphics[width=1in,height=1.25in,clip,keepaspectratio]{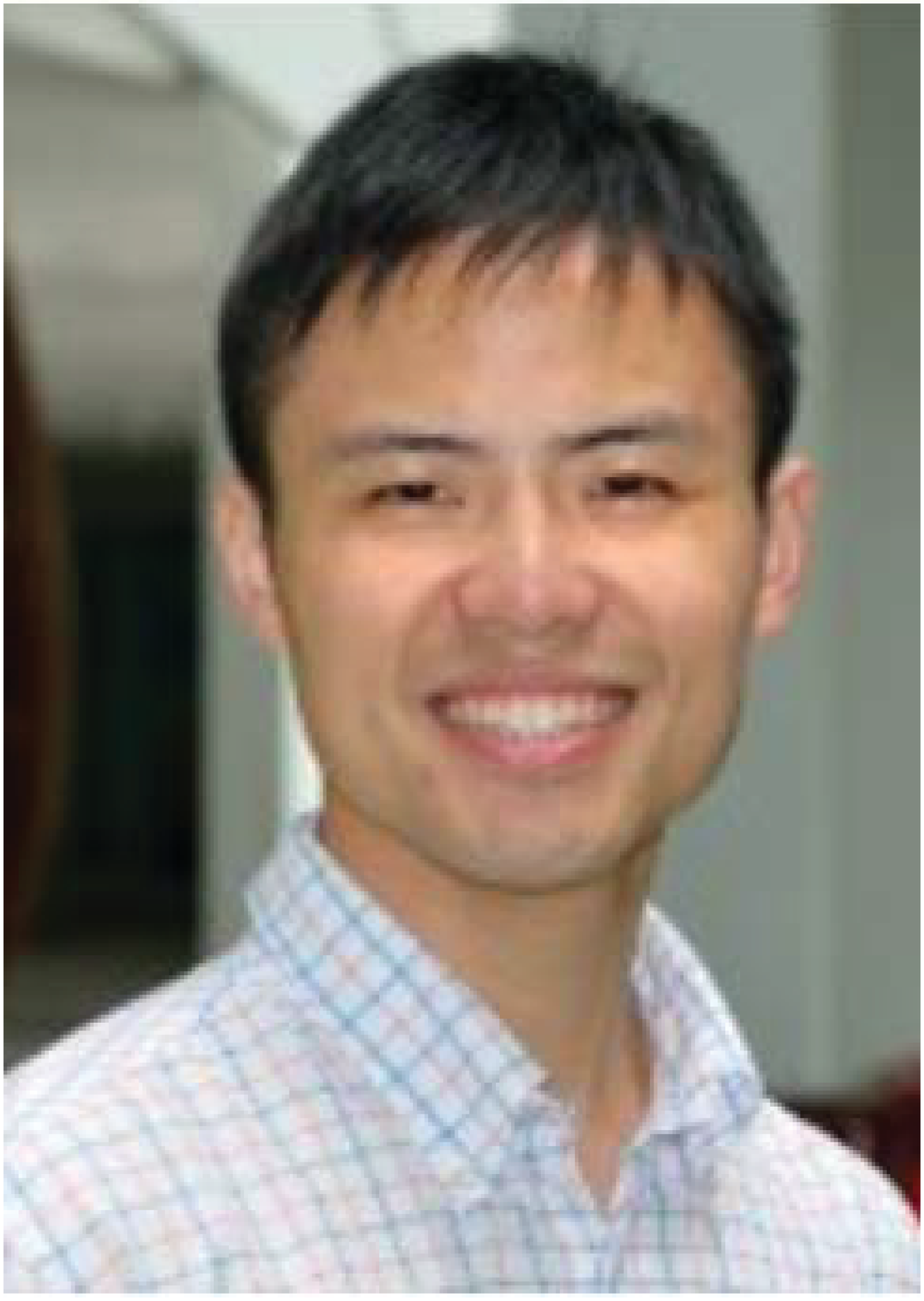}}] {Peng Cheng} (M'10) received the B.Sc. degree in automation and the Ph.D. degree in control science and engineering, from Zhejiang University, Hang Zhou, China, in 2004 and 2009, respectively. From 2012 to 2013, he worked as Research Fellow in Information System Technology and Design Pillar, Singapore University of Technology and Design. He is currently a Professor with the College of Control Science and Engineering, Zhejiang University, Hangzhou, China. His research interests include networked sensing and control, cyber-physical systems, and control system security.
\vspace{-0.85cm}
\end{IEEEbiography}
\begin{IEEEbiography}[{\includegraphics[width=1in,height=1.25in,clip,keepaspectratio]{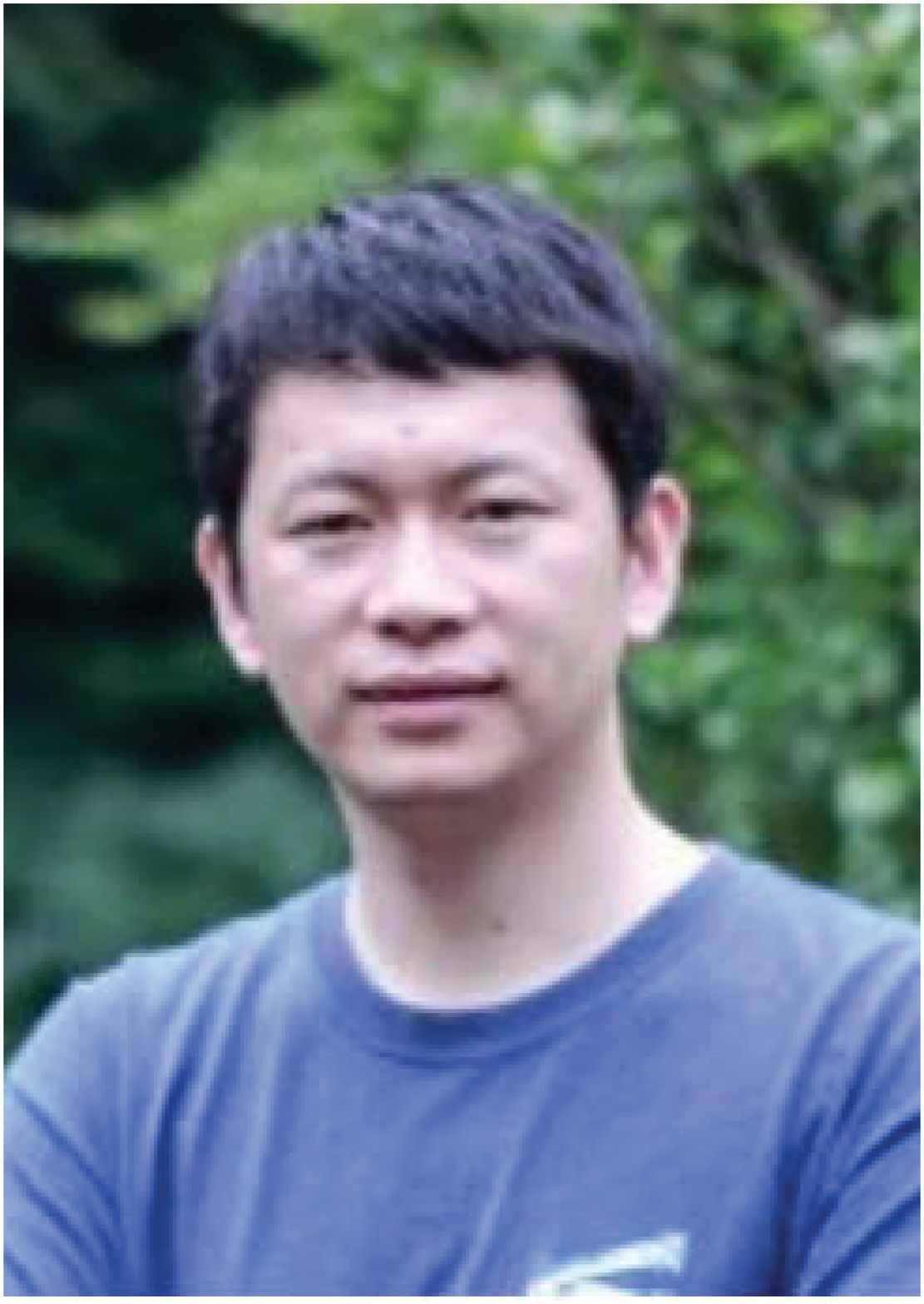}}] {Jiming Chen} (M'08-SM'11-F'18) received the B.Sc. and Ph.D. degrees in control science and engineering from Zhejiang University, Hangzhou, China, in 2000 and 2005, respectively. He was a Visiting Researcher with the University of Waterloo from 2008 to 2010. He is currently a Changjiang Scholars Chair Professor (MOE) with the College of Control Science and Engineering, the Deputy Director of the State Key Laboratory of Industrial Control Technology, and a member of the Academic Committee with Zhejiang University. His research interests include the Internet of Things, sensor networks, networked control, and control system security. He was a recipient of the Fok Ying Tung Young Teacher Award of the Ministry of Education and the IEEE ComSoc Asia-Pacific Outstanding Young Researcher Award. He is an IEEE VTS Distinguished Lecturer and IEEE Fellow.
\end{IEEEbiography}

\end{document}